\documentclass[journal=jacsat,manuscript=article]{achemso}

\usepackage{chemformula} % Formula subscripts using \ch{}
\usepackage[T1]{fontenc} % Use modern font encodings
\author{Sitaram Ramakrishnan}
\affiliation {Department of Quantum Matter, AdSE,
Hiroshima University, Higashi-Hiroshima 739-8530, Japan}
\email{niranj002@gmail.com}
\author{Jin-Ke Bao}
\affiliation {Laboratory of Crystallography,
University of Bayreuth, 95447 Bayreuth, Germany}
\alsoaffiliation[Second University]
{Department of Physics, Materials Genome Institute and International Center
for Quantum and Molecular Structures, Shanghai University,
 Shanghai 200444, People’s Republic of China}
\author{Claudio Eisele}
\affiliation {Laboratory of Crystallography,
University of Bayreuth, 95447 Bayreuth, Germany}
\author{Bikash Patra}
\affiliation{Department of Condensed Matter Physics and Materials Science,
Tata Institute of Fundamental Research, Mumbai 400005, India}
\author{Minoru Nohara}
\affiliation {Department of Quantum Matter, AdSE,
Hiroshima University, Higashi-Hiroshima 739-8530, Japan}
\author{Biplab Bag}
\affiliation{Department of Condensed Matter Physics
and Materials Science,
Tata Institute of Fundamental Research, Mumbai 400005, India}
\alsoaffiliation[Second University]
{Amity Institute of Applied Sciences,
Amity University Jharkhand, India‐834001}
\author{Leila Noohinejad}
\affiliation{P24, PETRA III, Deutsches Elektronen-Synchrotron DESY,
Notkestrasse 85, 22607 Hamburg, Germany}
\author{Martin Tolkiehn}
\affiliation{P24, PETRA III, Deutsches Elektronen-Synchrotron DESY,
Notkestrasse 85, 22607 Hamburg, Germany}
\author{Carsten Paulmann}
\affiliation{P24, PETRA III, Deutsches Elektronen-Synchrotron DESY,
Notkestrasse 85, 22607 Hamburg, Germany}
\alsoaffiliation[Second University]
{Mineralogisch-Petrographisches Institut,
Universit\"{a}t Hamburg, 20146 Hamburg, Germany}
\author{Achim M. Schaller}
\affiliation {Laboratory of Crystallography,
University of Bayreuth, 95447 Bayreuth, Germany}
\author{Toms Rekis}
\affiliation {Laboratory of Crystallography,
University of Bayreuth, 95447 Bayreuth, Germany}
\author{Surya Rohith Kotla}
\affiliation {Laboratory of Crystallography,
University of Bayreuth, 95447 Bayreuth, Germany}
\author{Andreas Sch\"{o}nleber}
\affiliation {Laboratory of Crystallography,
University of Bayreuth, 95447 Bayreuth, Germany}
\author{Arumugam Thamizhavel}
\affiliation{Department of Condensed Matter Physics and Materials Science,
Tata Institute of Fundamental Research, Mumbai 400005, India}
\author{Bahadur Singh}
\affiliation{Department of Condensed Matter Physics and Materials Science,
Tata Institute of Fundamental Research, Mumbai 400005, India}
\email{bahadur.singh@tifr.res.in}
\author{Srinivasan Ramakrishnan}
\affiliation{Department of Condensed Matter Physics and Materials Science,
Tata Institute of Fundamental Research, Mumbai 400005, India}
\alsoaffiliation[Second University]
{IISER, Dr. Homi Bhabha Road, Pashon, Pune-411008, India}
\email{ramky07@gmail.com}
\author{Sander van Smaalen}
\affiliation{Laboratory of Crystallography,
University of Bayreuth, 95447 Bayreuth, Germany}
\email{smash@uni-bayreuth.de}

\title{Coupling between colossal charge density wave
ordering and magnetism in Ho$_2$Ir$_3$Si$_5$}

\begin{document}

%\clearpage

\begin{tocentry}

\includegraphics{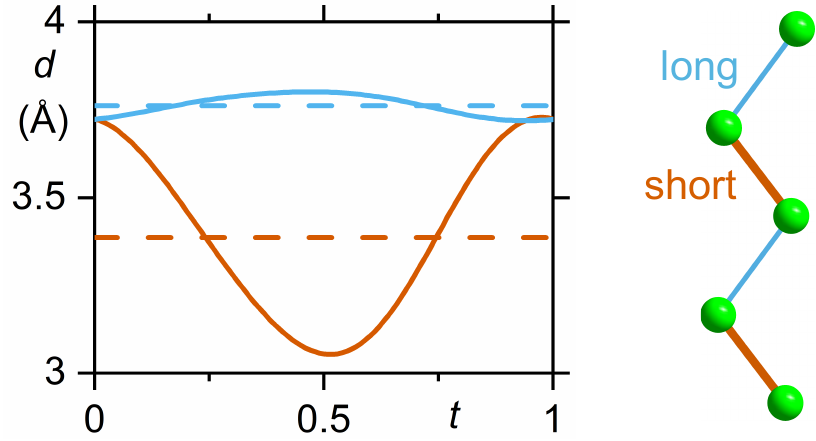}

%Some journals require a graphical entry for the Table of Contents.
%This should be laid out ``print ready'' so that the sizing of the

%Inside the \texttt{tocentry} environment, the font used is Helvetica
%8\,pt, as required by \emph{Journal of the American Chemical

%The surrounding frame is 9\,cm by 3.5\,cm, which is the maximum
%permitted for  \emph{Journal of the American Chemical Society}
%graphical table of content entries. The box will not resize if the
%content is too big: instead it will overflow the edge of the box.

%This box and the associated title will always be printed on a
%separate page at the end of the document.

\end{tocentry}

\begin{abstract}
Ho$_2$Ir$_3$Si$_5$ belongs to the family of
three-dimensional (3D) $R_2$Ir$_3$Si$_5$ ($R$ = Lu, Er and Ho)
compounds that exhibit a colossal first-order
charge density wave (CDW) transition where there is
a strong orthorhombic-to-triclinic distortion of the
lattice accompanied by superlattice reflections.
The analysis by single-crystal X-ray diffraction
(SXRD) has revealed that the Ir-Ir zigzag
chains along \textbf{c} are responsible for the CDW
in all three compounds.
The replacement of the rare earth element from
non-magnetic Lu to magnetic Er or Ho
lowers $T_{CDW}$, where
$T_{CDW Lu}$ = 200 K, $T_{CDW Er}$ = 150 K
and $T_{CDW Ho}$ = 90 K.
Out of the three compounds, Ho$_2$Ir$_3$Si$_5$ is
the only system where
second-order superlattice reflections could be observed,
indicative of an anharmonic shape of the modulation wave.
The CDW transition is observed as anomalies in the
temperature dependencies of the specific heat,
electrical conductivity and magnetic susceptibility,
which includes a large hysteresis of 90 to 130 K for
all measured properties,
thus corroborating the SXRD measurements.
Similar to previously reported Er$_2$Ir$_3$Si$_5$,
there appears to be a coupling between CDW and
magnetism such that the Ho$^{3+}$
magnetic moments are influenced by the CDW
transition, even in the paramagnetic state.
Moreover, earlier investigations on polycrystalline
material revealed antiferromagnetic (AFM) ordering at
$T_N$ = 5.1 K, whereas AFM order is suppressed and
only the CDW is present in our highly ordered single-crystal.
First-principles calculations predict
Ho$_2$Ir$_3$Si$_5$ to be a
metal with coexisting electron and hole pockets
at the Fermi level.
The Ho and Ir atoms have spherically symmetric
metallic-type charge density distributions
that are prone to CDW distortion.
Phonon calculations affirm that the Ir
atoms are primarily responsible for the CDW
distortion, which is in agreement with the experiment.
\end{abstract}


\clearpage

\section{\label{ho235_s_introduction}%
Introduction}

Structural phase transitions, such as the development
of charge density waves (CDWs),
%where there is a rearrangement of atoms due to lowering
%of symmetry and presence of superlattice reflections
have attracted continued interest from condensed-matter
physicists and chemists.
Originally, the manifestation of a CDW was proposed
to originate in Fermi surface nesting (FSN), as it is
present in quasi-one-dimensional metals (1D) \cite{gruener1994a}.
More recently, alternative mechanisms were proposed that
explain, for example, the formation of CDWs
by $q$-dependent electron-phonon coupling (EPC)
in three-dimensional (3D) CDW compounds.
Examples of 3D metals with CDWs include
$\alpha$-Uranium \cite{marmeggijc1982a},
CuV$_2$S$_4$ \cite{flemingrm1981a, kawaguchi2012a, okadah2004a, ramakrishnan2019a},
La$_3$Co$_4$Sn$_{13}$ \cite{slebarski2013a, otomo2016a, welsch2019a},
$R$Te$_3$ ($R$ = La, Sm, Gd, Tb, Dy, Ho, Er, Tm)
\cite{dimasie1995a, run2008a, banerjee2013a},
$R$Te$_2$ ($R$ = La, Ce) \cite{dimasie1996a, shim2004a},
$R$$_5$Ir$_4$Si$_{10}$ ($R$ = Dy, Ho, Er, Yb, Lu) \cite{ramakrishnan2017a},
Sm$_2$Ru$_3$Ge$_5$ \cite{kuo2020a},
EuAl$_4$ \cite{nakamura2015a, shimomura2019a, kaneko2021a, ramakrishnan2022a,meierwr2022a}
and CuIr$_{2-x}$Cr$_x$Te$_4$ \cite{zeng2022a}.
In several of these compounds a coexistence and
competition exists between the CDW and
antiferromagnetic (AFM) order or superconductivity (SC).
In the family of compounds $R$NiC$_2$
($R$ = Ce, Pr, Nd, Sm, Gd, Tb, Dy, Ho, Er, Tm)
\cite{romanm2018a,shimomura2009a,wolfel2010a,shimomuras2016a, kolinciokk2017a,maeda2019a,kolincio2020a},
the CDW competes with AFM.
In the case of SmNiC$_2$ ferromagnetic (FM) order
completely destroys the CDW \cite{shimomura2009a, wolfel2010a}.
On the other hand, Kolincio \textit{et al.} \cite{kolincio2020a}
have established the coexistence of a CDW
and field-induced FM order in TmNiC$_2$,
suggesting strong coupling of the rare-earth spins to the CDW
in these rare-earth materials.
However, the magnetic susceptibility of the paramagnetic
regime does not exhibit anomalies at the CDW transitions.

Er$_2$Ir$_3$Si$_5$ differs from most magnetic CDW compounds,
in that the magnetic susceptibility of the paramagnetic
state exhibits an anomaly at the CDW transition
\cite{ramakrishnan2020a}.
Presently, we report a similar effect for Ho$_2$Ir$_3$Si$_5$.
On the other hand, the CDW of non-magnetic Lu$_2$Ir$_3$Si$_5$
coexists with SC at low temperatures \cite{sangeetha2015a}.

Investigations of the CDW in Lu$_2$Ir$_3$Si$_5$
has been extensively done
\cite{singhy2004a, singhy2005a, kuoyk2006a, leemh2011a}.
Studies by transmission electron microscopy (TEM)
reported the modulation wave vector as
\textbf{q} = $\sigma(\bar{1}21)$ with $\sigma$ = 0.23--0.25
around 200 K \cite{leemh2011a}.
Recently, the modulated crystal structure of Lu$_2$Ir$_3$Si$_5$
was reported as similar to that of
Er$_2$Ir$_3$Si$_5$ \cite{ramakrishnan2021a}.
In both systems it was elucidated that the CDW resides
on the zigzag chains of Ir atoms along \textbf{c}.
Although the rare earth element
%plays a less significant role in the formation
is not directly involved in the stabilization
of the CDW,
we have earlier proposed that it would indirectly
influence the CDW through its size (atomic radius).
Er has a larger atomic radius than to Lu has
(2.26 vs 2.17 \AA{}),\cite{clementi1967a}
while the transition occurs 50 K lower in
Er$_2$Ir$_3$Si$_5$ than in Lu$_2$Ir$_3$Si$_5$.
Here, we find that there is no such simple relationship
between atomic radius and $T_{CDW}$.
The atomic radii of Ho and Er are equal,
but $T_{CDW}$ = 90 K of Ho$_2$Ir$_3$Si$_5$
is much lower than $T_{CDW}$ = 150 K of
Er$_2$Ir$_3$Si$_5$.
%
%and it is more
%sluggish than compared to the transitions of
%Er$_2$Ir$_3$Si$_5$ and
%Lu$_2$Ir$_3$Si$_5$, which is anomalous.

From SXRD we found that Ho$_2$Ir$_3$Si$_5$
undergoes a large distortion of the lattice where
the symmetry is lowered from orthorhombic ($Ibam$)
to triclinic ($I\bar{1}$) below the transition
temperature $T_{CDW}$.
Simultaneously, superlattice reflections
appear at incommensurate positions, with values of
$\mathbf{q}$ = $[0.2494(2),\: 0.4978(2),\: 0.2488(2)]$ at 70 K.
The CDW is supported by zigzag chains of Ir atoms
in all three compounds $R_2$Ir$_3$Si$_5$.
Ho$_2$Ir$_3$Si$_5$ differs from Lu$_2$Ir$_3$Si$_5$
and Er$_2$Ir$_3$Si$_5$ by the presence of
second-order superlattice reflections in its SXRD,
indicative of anharmonic contributions to
the displacive modulation wave.

We present for Ho$_2$Ir$_3$Si$_5$ the temperature
dependencies of the electrical resistivity $\rho(T)$,
the specific heat $C_p(T)$ and the magnetic
susceptibility $\chi(T)$.
They show anomalies in agreement with a first-order
CDW transition with a hysteresis of about 40 K.
Due to the strain imposed by the transition onto
the crystal,
it cracks which can be seen clearly from the
resistivity measurements.
Such cracking of crystals has also been observed
in other CDW compounds, like BaFe$_2$Al$_9$ \cite{meierwr2021a}.
From the magnetic susceptibility we observed that
there is an influence of the CDW on the effective
magnetic moment of Ho$^{3+}$, which
is similar to what was reported in  Er$_2$Ir$_3$Si$_5$ \cite{ramakrishnan2020a}.
Such an effect of the CDW on the rare earth spin
in the paramagnetic state
is not commonly seen in rare earth compounds.
Earlier studies have shown that the disorder
in the polycrystalline material
allows long-range AFM order to develop at
low temperatures \cite{singhy2004a, singhy2005a}.
However, in the present high-quality
single crystal, long-range AFM order is absent
and only the CDW is observed.
Here, we elucidate the differences and similarities
in the incommensurate
structure of  Ho$_2$Ir$_3$Si$_5$ by comparing
with Lu$_2$Ir$_3$Si$_5$ and Er$_2$Ir$_3$Si$_5$.
Furthermore, we explore the coupling of the CDW
and magnetism of Ho$^{3+}$ spins and absence of
long-range AFM order.

\section{\label{ho235_s_experiment_computation}%
Experimental and computational details}

\subsection{\label{ho235_s_crystal_growth}%
Crystal growth and characterization}

\begin{figure}%[H]
\includegraphics[width=80mm]{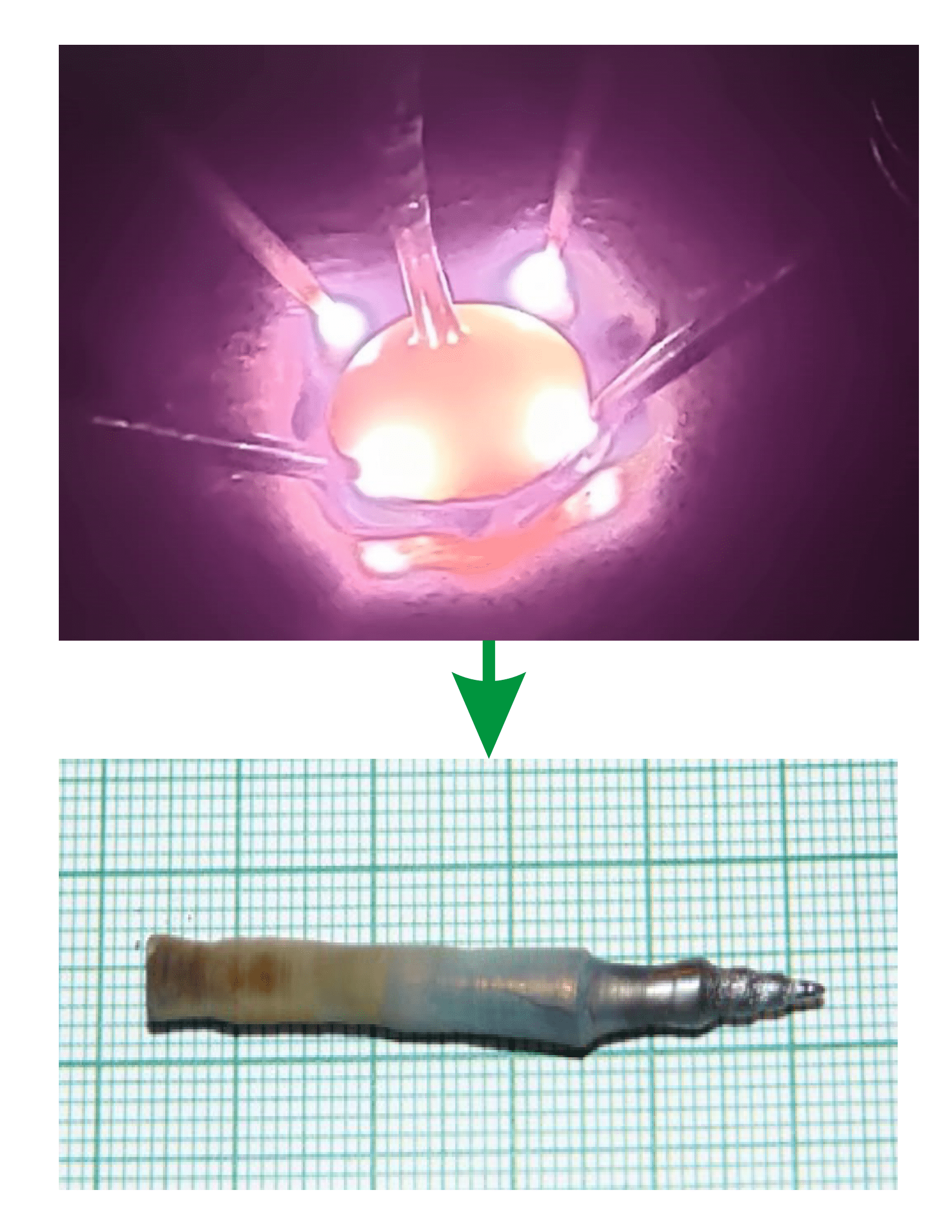}
\caption{Crystal growth in a tetra arc furnace
by the Czochralski method.
A bar was obtained of about 50 mm length
and 15 mm in diameter.}
\label{fig:ho2ir3si5_crystalgrowth}
\end{figure}

A single crystal of Ho$_2$Ir$_3$Si$_5$ has been grown by the Czochralski method in a tetra-arc furnace
as shown in Fig. \ref{fig:ho2ir3si5_crystalgrowth}.
We have chosen the Czochralski over flux growth or any other technique because
the individual elements have high melting points.  Furthermore, iridium metal does not dissolve in most
of the typical fluxes.
To start with, high purity
individual elements of Ho : Ir : Si (99.99\% for Ho and Ir, and 99.999\% for Si) were taken in the stoichiometric ratio 2 : 3 : 5, amounting to about 8 to 9 g,
and melted repeatedly to ensure its homogeneity. Next, a seed crystal was cut from this polycrystalline
ingot for the purpose of crystal growth.  The polycrystalline seed was gently inserted into the molten solution
and initially pulled at a rapid speed of about 80 mm/h.  The temperature of the melt was adjusted such
that a necking is formed and then we employed a pulling speed of about 10 mm/h throughout the growth process.
About 50 mm long ingot was pulled.
%The grown ingot was checked for the
%Laue diffraction to confirm the single
%crystallinity  and to align it along the
%principal crystallographic directions.
Energy-dispersive X-ray spectroscopy (EDX) was
used to verify the chemical composition.

\subsection{\label{ho235_s_sxrd}%
Single-crystal X-ray diffraction (SXRD):
data collection and data processing.}

\begin{table}%[H]
\caption{\label{tab:ho2ir3si5_cdw_crystalinfo}%
Crystallographic data of crystal A of
Ho$_2$Ir$_3$Si$_5$ at 200 K (periodic phase) and 70 K (CDW phase).}
%\scriptsize
\small
\centering
\begin{tabular}{ccc}
\hline
Temperature (K) & 200 & 70  \\
Crystal system & Orthorhombic& Triclinic  \\
Space/Superspace group & $Ibam$  &
$I\bar{1}(\sigma_1\:\sigma_2\:\sigma_3)0$ \\
Space/Superspace group No. \cite{stokesht2011a} & 72 & {2.1.1.1}   \\
$a$ (\AA{}) &9.9023(4) &9.8356(5)  \\
$b$ (\AA{}) &11.3747(3)    &11.4902(4)     \\
$c$ (\AA{}) &5.7745(3)    &5.7304(3)  \\
$\alpha$ (deg) & 90 & 89.983(3) \\
$\beta$ (deg) & 90 & 91.772(2) \\
$\gamma$ (deg) &   90 &  89.975(1)  \\
Volume (\AA{}$^3$) & 650.42(5) & 647.32(5)  \\
Wave vector \textbf{q} & -
& (0.2494(2), 0.4978(2), 0.2488(2))  \\
$Z$ & 4 & 4 \\
Wavelength (\AA{}) & 0.50000 &0.50000  \\
Detector distance (mm) &110 &110  \\
$2\theta$-offset (deg) &0 &0  \\
$\chi$-offset (deg) &-60 & -60  \\
Rotation per image (deg) & 1 & 1  \\
$(\sin(\theta)/\lambda)_{max}$ (\AA{}$^{-1}$) &0.683589& 0.684039 \\
Absorption, $\mu$ (mm$^{-1}$) & 34.823 & 34.990  \\
T$_{min}$, T$_{max}$ & 0.0220, 0.0501 & 0.0230, 0.0501  \\
Criterion of observability & $I>1.5\sigma(I)$ & $I>1.5\sigma(I)$ \\
Number of $(m = 0)$ reflections \\
measured &  3765  & 2706  \\
unique (obs/all) & 451/470 &1298/1474  \\
Number of $(m = 1)$ reflections \\
measured & - & 12770  \\
unique (obs/all) & - & 2779/6454  \\
Number of $(m = 2)$ reflections \\
measured & -& 12725 \\
unique (obs/all) & -& 714/3263  \\
$R_{int}$ $(m = 0)$ (obs/all) &0.0487/0.0487 &0.0287/0.0288  \\
$R_{int}$ $(m = 1)$ (obs/all) &- &0.0828/0.1051 \\
$R_{int}$ $(m = 2)$ (obs/all) &- &0.0967/0.1833 \\
No. of parameters &31 &147 \\
$R_{F }$ $(m = 0)$  (obs) &0.0296 &0.0578 \\
$R_{F }$ $(m = 1)$ (obs) &- &0.0798 \\
$R_{F }$ $(m = 2)$ (obs) &- &0.2148  \\
$wR_{F }$ $(m = 0)$ (all) &0.0380 &0.0717 \\
$wR_{F }$ $(m = 1)$ (all) &- &0.0973 \\
$wR_{F }$ $(m = 2)$ (all) &- &0.3748 \\
$wR_{F }$ all (all) &0.0380 &0.0879 \\
GoF (obs/all) &1.83/1.33 &1.69/1.23 \\
$\Delta\rho_{min}$, $\Delta\rho_{max}$ (e \AA{}$^{-3}$) &
 -3.00, 2.94 &-14.56, 14.53  \\
 \hline
\end{tabular}
\end{table}

Small pieces of single-crystalline Ho$_2$Ir$_3$Si$_5$ were obtained
by crushing a large single crystal, from which crystal A of dimensions
0.15$\times$0.07$\times$0.1 mm${^3}$ was selected for a single-crystal X-ray
diffraction (SXRD) experiment at beamline P24 of PETRA-III Extension at DESY in Hamburg,
Germany.
SXRD was measured at station EH2 of beamline P24,
employing radiation of a wavelength
of $\lambda_{P24}$ = 0.50000 \AA{}.
For further details regarding data
collection refer to the supporting
information \cite{ho2ir3si5suppmat2022a}.

The EVAL15 software suite \cite{schreursamm2010a}
was used for processing the SXRD data.
SADABS \cite{sheldrick2008}  was used for scaling
and absorption correction, with Laue symmetry $mmm$
for the data in the periodic phase
and $\bar{1}$ for the CDW phase.
As the crystal structure in the CDW phase is
incommensurately modulated we had to use
the superspace approach \cite{van2007incommensurate, wagner2009a, stokesht2011a} to index and integrate the data.
Section S2 in the supporting
information \cite{ho2ir3si5suppmat2022a}
provides further details.
The resulting reflection file was imported
into JANA2006 \cite{petricekv2016a, petricekv2014a}.
Table \ref{tab:ho2ir3si5_cdw_crystalinfo} gives
the crystallographic information at 200 K (periodic phase)
and at 70 K (incommensurate phase).
The crystallographic data at other temperatures
and details regarding SXRD data processing
are given in the Supporting Information.\cite{ho2ir3si5suppmat2022a}

\subsection{\label{sec:ho2ir3si5_physical_properties}%
Physical properties}

A commercial superconducting quantum interference device (SQUID) magnetometer (MPMS5,
Quantum Design, USA) was used to measure dc magnetic susceptibility in a field
of 100 mT as a function of temperature from 2 to
300 K. The electrical resistivity between 1.8 and 300 K was
measured by the standard dc four probe technique on
a commercial physical property measurement system (PPMS, Quantum
Design, USA). The specific heat data were measured both
on PPMS as well as using a commercial differential scanning calorimeter (DSC)
setup.

\subsection{\label{sec:ho2ir3si5_dft}%
Density functional theory calculations}

Density functional theory (DFT) based calculations
were performed using the projector augmented wave
(PAW)~\cite{blochl1994projector} method
as implemented in the Vienna \textit{ab initio}
simulation package (VASP).\cite{kresse1996efficient}
Exchange-correlation effects were included
using the Perdew-Burke-Ernzerhof
(PBE)~\cite{perdew1996generalized} version
of the generalized gradient approximation.
A $8\times 10 \times 12$ $\Gamma$-centered $k$-mesh
was used for the Brillouin zone (BZ) sampling with
an energy cut-off of $380$ eV for
the plane-wave basis set.
Spin-orbit coupling (SOC) effects were taken into
account self-consistently to consider the
relativistic effects.
We employed Ho$^{3+}$ potential by considering
the remaining 4\textit{f} electrons as core electrons.
Experimental lattice parameters were used, while
the internal atomic positions were relaxed
until the residual forces on each atom were
less than 0.0001 eV/\AA{}.
The VESTA\cite{momma2008vesta} program was used
to visualize the charge density distributions.
The phonon spectrum was obtained using a
$2\times 2\times 2$ supercell following the
frozen phonon method as implemented in
the phonopy~\cite{togo2015first} package.

\section{\label{sec:ho2ir3si5_results_discussion}%
Results and discussion}

\subsection{\label{sec:ho2ir3si5_cdw_structure}%
Analysis of the CDW structure}

Unlike Er$_2$Ir$_3$Si$_5$,\cite{ramakrishnan2020a}
the manifestation of the CDW appears to be
sluggish in Ho$_2$Ir$_3$Si$_5$.
The crystal was initially cooled from room
temperature down to 70 K, where
it remained in the orthorhombic phase,
despite $T_{CDW}$ being about 90 K
according to the physical property measurements.

\begin{figure}%[H]
\includegraphics[width=80mm]{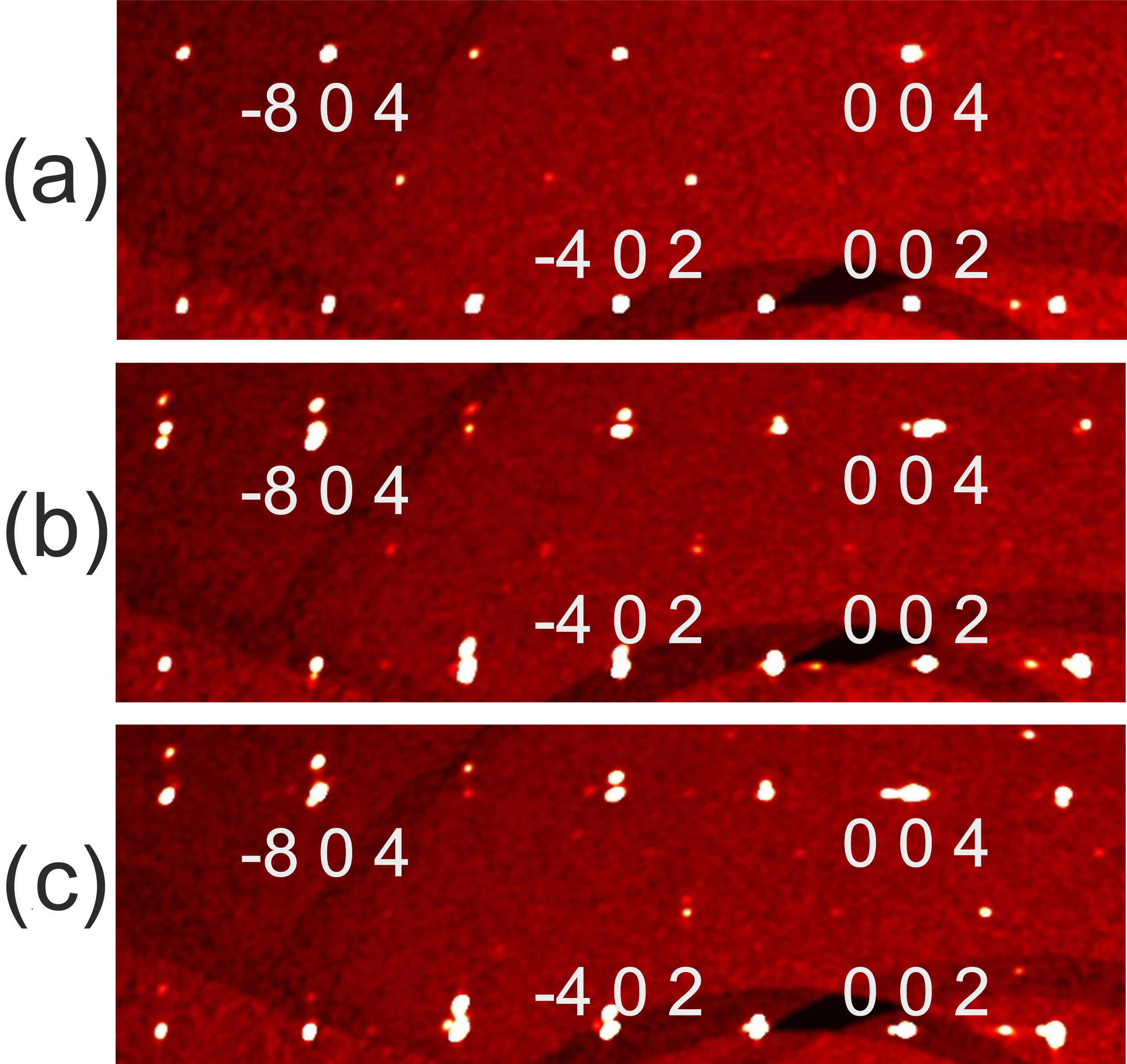}
\vspace{2mm}\\
\includegraphics[width=100mm]{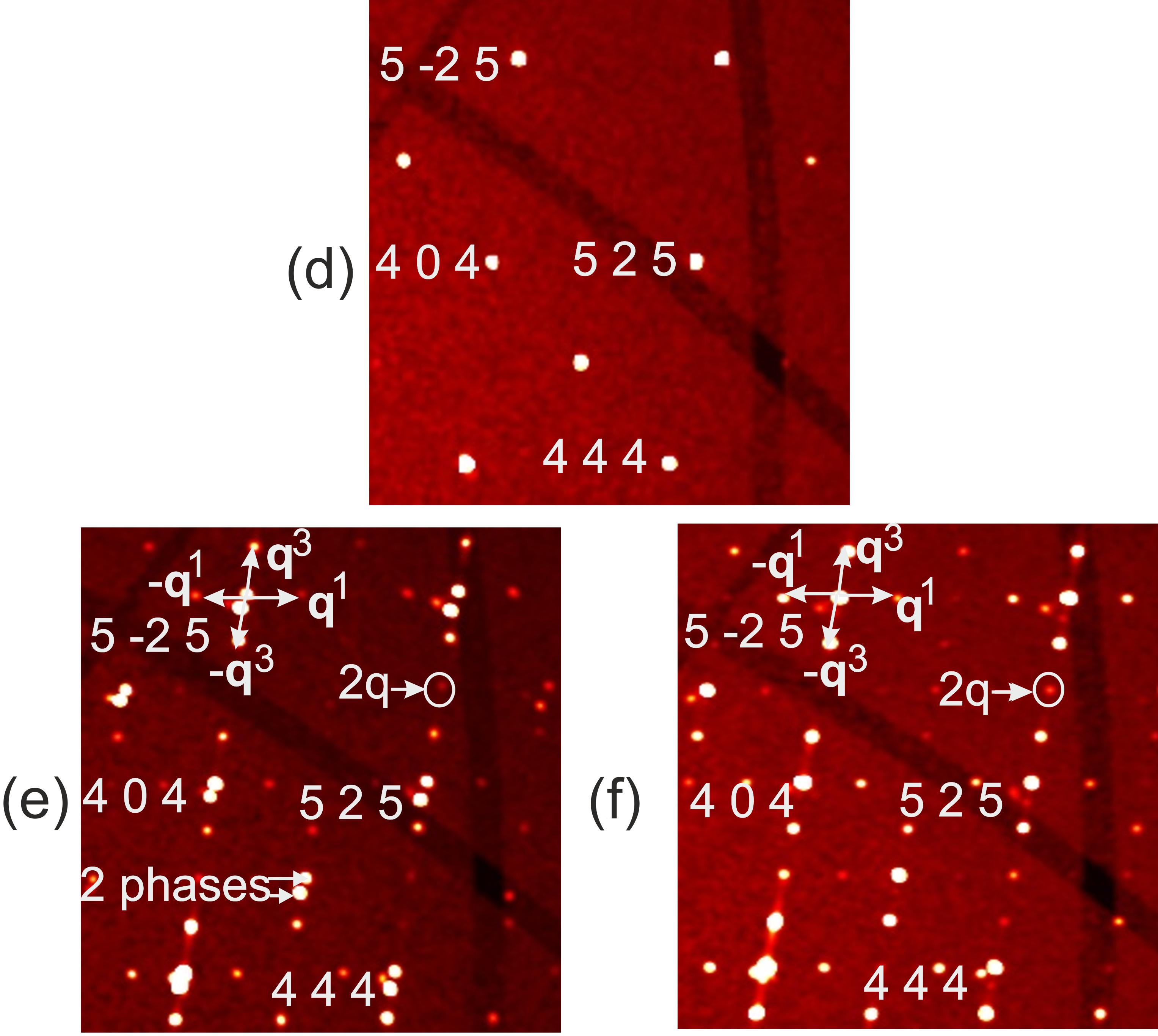}
\caption{\label{fig:ho2ir3si5_unwarp}%
(a), (b), (c) The $h\,0\,l$ plane, and
(d), (e), (f) the $h\,k\,h$ plane reconstructed
from measured SXRD data.
(a), (d) are for SXRD on the periodic phase at 200 K.
(b), (e) show the intermediate phase (undercooled state),
where threefold splitting of the reflections can be observed at 20 K,
indicating coexistence of CDW and periodic phases.
(c), (f) are for SXRD on the CDW phase at 50 K, showing
groups of two instead of groups of three split reflections.
The degree of splitting increases with the value of $h$.
Panels (e) and (f) show satellite reflections of order
$m = 2$.}
\end{figure}

\begin{figure}%[H]
\includegraphics[width=80mm]{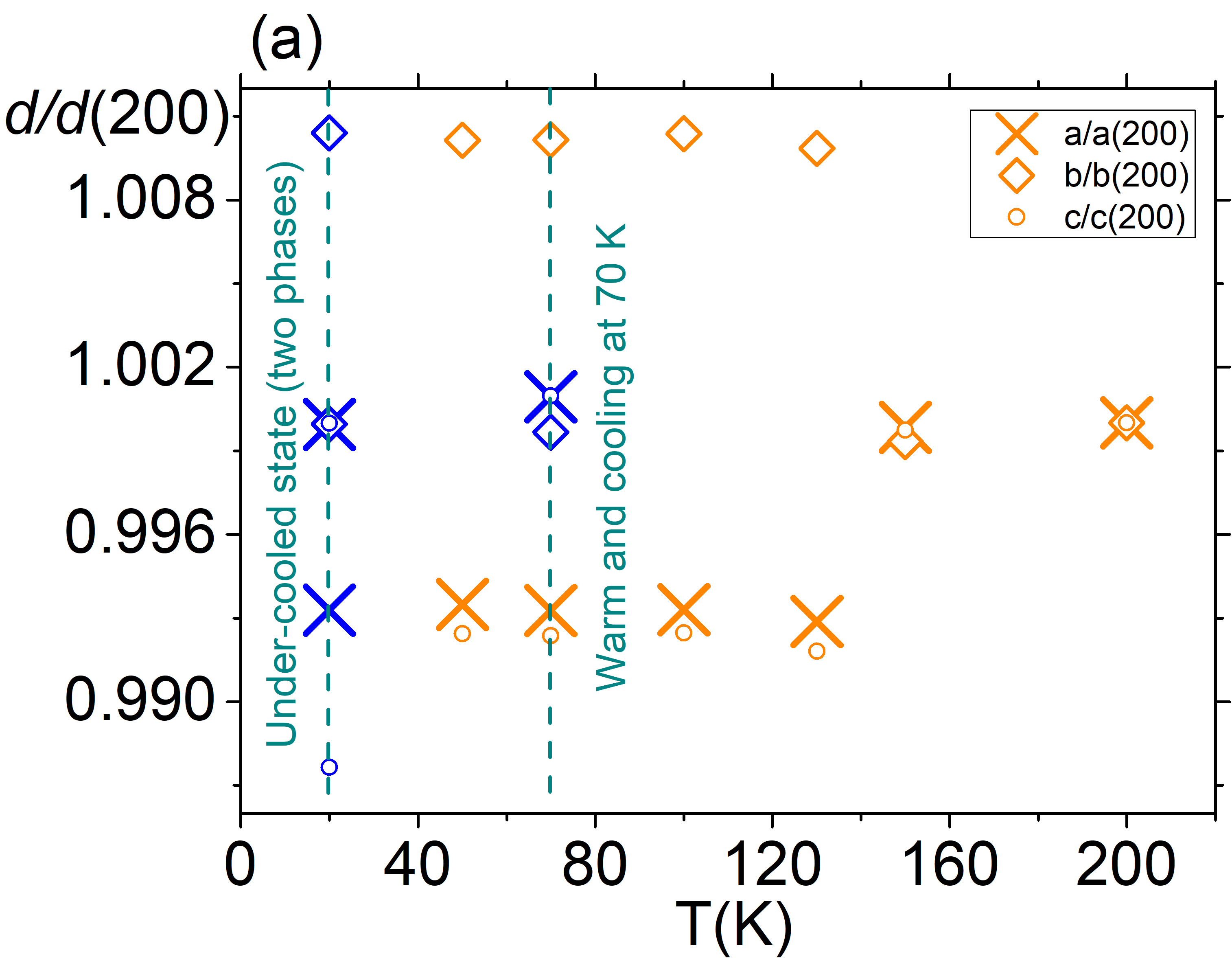}
\includegraphics[width=80mm]{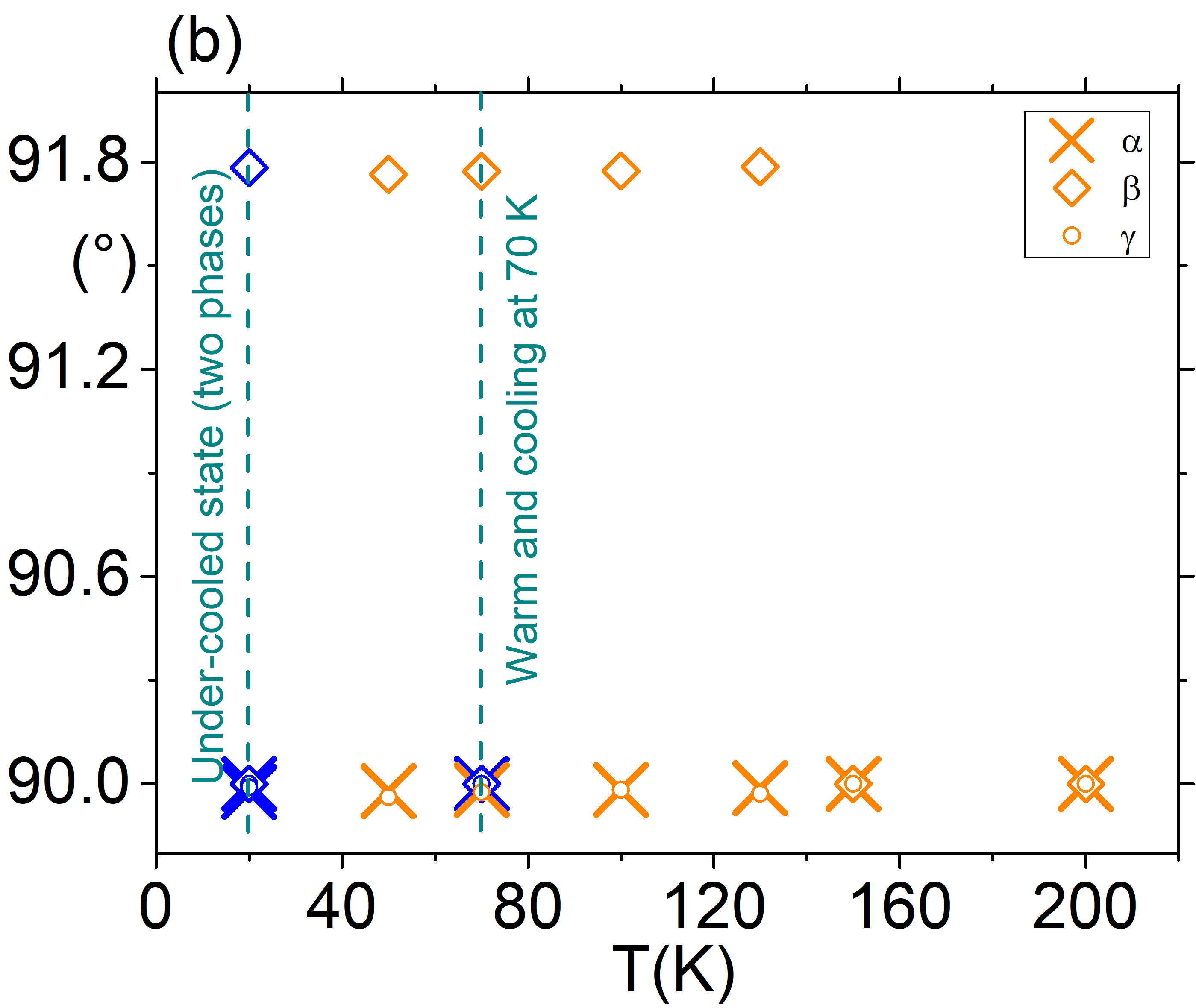}
\\
\includegraphics[width=80mm]{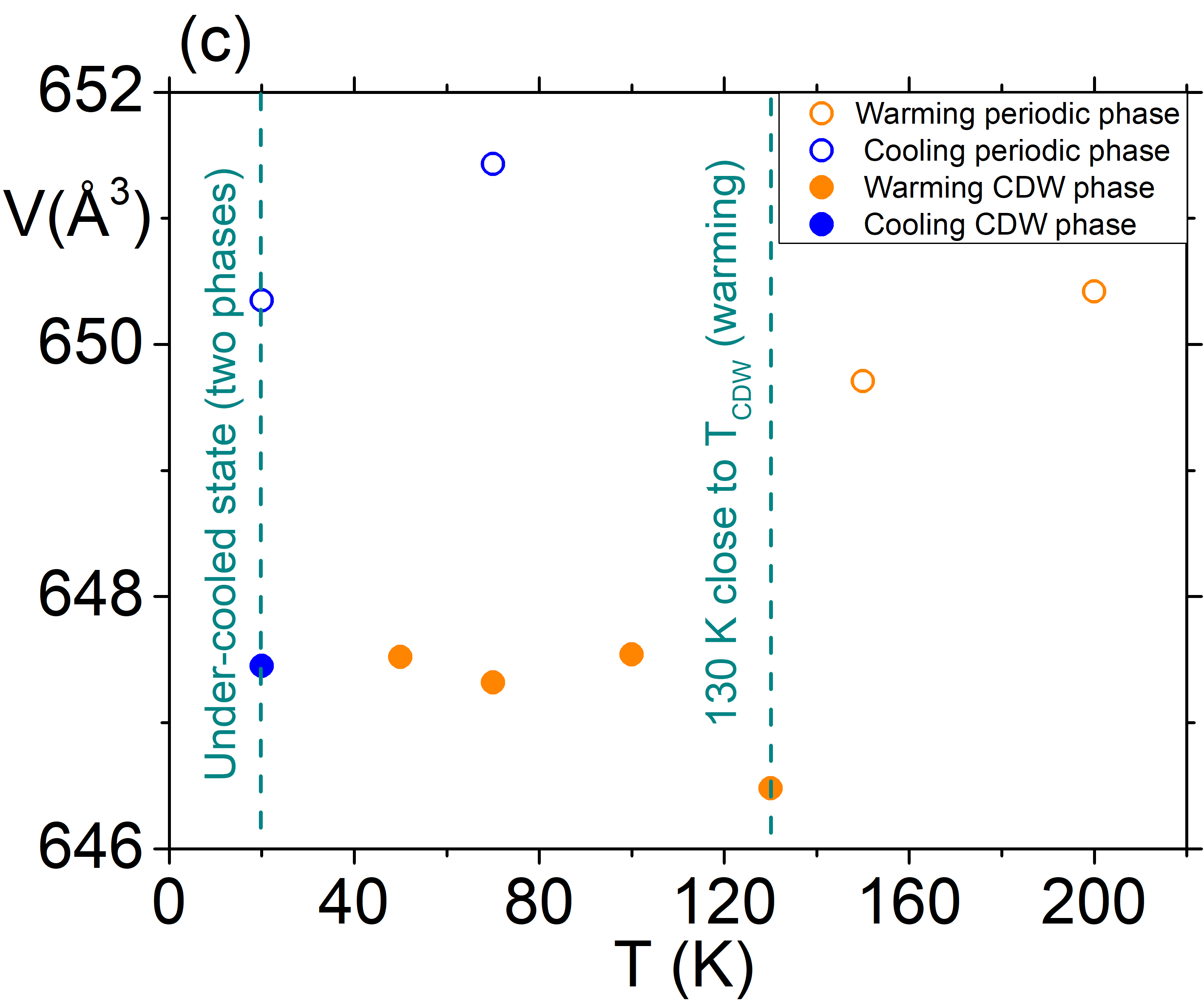}
\includegraphics[width=80mm]{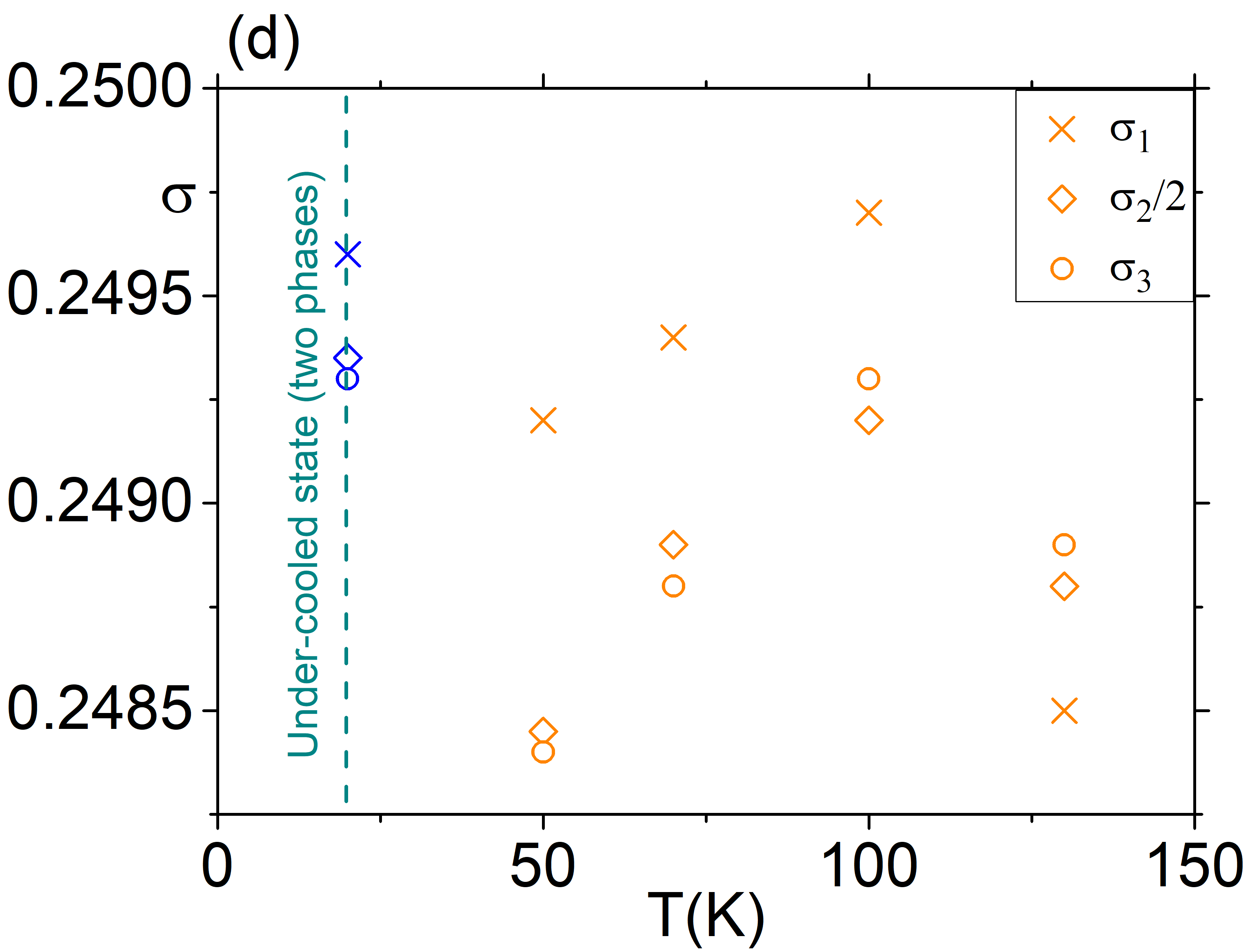}
\caption{\label{fig:ho2ir3si5_lattice}%
Temperature dependence of
(a) the lattice parameters $a$, $b$ and $c$
relative to their values at $T = 200$ K:
$a(200) = 9.9023(4)$, $b(200) = 11.3747(3)$
and $c(200) = 5.7745(3)$ \AA{};
(b) the lattice parameters $\alpha$, $\beta$
and $\gamma$;
(c) the volume of the unit cell;
and
(d) components of the modulation wave vector \textbf{q}.
Orange and blue symbols refer to data obtained
during heating and cooling of the crystal, respectively.
The undercooled state at 20 K has coexisting
CDW and periodic phases.}
\end{figure}

It is possible that the crystal was cooled too
rapidly, thereby not allowing it to settle
in the CDW phase, and resulting in an undercooled state.
Upon further lowering of the temperature to 20 K,
we observed the coexistence of orthorhombic and
CDW phases, indicative
of a first-order transition
(Figs. \ref{fig:ho2ir3si5_unwarp}
and \ref{fig:ho2ir3si5_lattice}).
The CDW phase is characterized by superlattice
reflections at positions
$\mathbf{q}$ = $[0.2496(3),\: 0.4987(3),\: 0.2493(3)]$,
accompanied by a large monoclinic distortion
of the lattice ($\beta = 91.784(3)$ deg),
similar to $R_2$Ir$_3$Si$_5$ ($R$ = Lu, Er)
(details are given in the Supporting
Information\cite{ho2ir3si5suppmat2022a}).

The severity of the distortions induces a colossal
strain in the crystal, causing the diffraction spots
in SXRD to become
broad and elongated [Figs. \ref{fig:ho2ir3si5_unwarp}(b)
and \ref{fig:ho2ir3si5_unwarp}(c)],
resulting in many partially overlapped
reflections which cannot be used for structural analysis.
Sometimes the strain is too much for the
crystal to handle,
such that it physically shatters, rendering
it unusable for further investigations.
Such destructive behaviour of the CDW has also
been reported for BaFe$_2$Al$_9$ \cite{meierwr2021a}.

Another important difference
to Lu$_2$Ir$_3$Si$_5$ and Er$_2$Ir$_3$Si$_5$
is that we observe in SXRD on Ho$_2$Ir$_3$Si$_5$
second-order superlattice reflections $(m = 2)$
in addition to first-order superlattice
reflections $(m = 1)$
[Figs. \ref{fig:ho2ir3si5_unwarp}(e) and
\ref{fig:ho2ir3si5_unwarp}(f)].
The orthorhombic phase disappears after the
crystal is heated to 50 K, and only the CDW
phase remains at temperatures 50--130 K
(Figs. \ref{fig:ho2ir3si5_unwarp}(c) and
\ref{fig:ho2ir3si5_lattice}).
Further heating to 150 and 200 K causes the crystal to
enter the orthorhombic phase and the CDW phase disappears
[Fig. \ref{fig:ho2ir3si5_unwarp}(a)].
Similar to Lu$_2$Ir$_3$Si$_5$ and Er$_2$Ir$_3$Si$_5$,
there is a lowering
of the point symmetry from orthorhombic
to triclinic which causes the crystal to be
twinned with four orientations \cite{parson2003a}.

Figure \ref{fig:ho2ir3si5_lattice} shows the
temperature dependence of the lattice parameters.
One can observe the clear distortion of the lattice
upon entering the CDW state,
including an expansion of $b$ and contractions
of $a$ and $c$.
Furthermore, the change in lattice type at $T_{CDW}$
appears discontinuous, in agreement with the
first-order character of the phase transition.
Barring 20 K and 130 K as 20 K is an undercooled
mixed state and 130 K is on the onset of $T_{CDW}$
warming and therefore unreliable, it can be inferred
from  Fig. \ref{fig:ho2ir3si5_lattice}(d)
that the modulation wave vector \textbf{q} decreases
with temperature.

The presence of second-order satellite reflections
required refinement of the crystal structure with
higher order harmonics for the modulation functions.
Table \ref{tab:ho2ir3si5_refmod_compare}
provides a comparison of the models to see which
provides the best
fit to the SXRD data in the CDW phase.

\begin{table}%%[H]
\caption{\label{tab:ho2ir3si5_refmod_compare}%
Quality of the fit to the SXRD data at 70 K
for three structure models for the CDW phase.
See text for models A, B and C.
The number of unique reflections is
1298/1474 for obs/all main reflections $(m = 0)$,
2779/6454 for obs/all $m = 1$ satellites,
and
714/3263 for obs/all $m = 2$ satellites.
Criterion of observability is $I>1.5\sigma(I)$.}
\centering
\begin{tabular}{cccc}
\hline
Models  & A  &  B & C \\
No. of parameters &147 &177 &177 \\
$R_{F }$ $(m = 0)$  (obs) &0.0578 &0.0577 &0.0535 \\
$R_{F }$ $(m = 1)$ (obs) &0.0798 &0.0796 &0.0672 \\
$R_{F }$ $(m = 2)$ (obs) &0.2148 &0.2119 &0.0894 \\
$wR_{F }$ $(m = 0)$ (all) &0.0717 &0.0717 &0.0695 \\
$wR_{F }$ $(m = 1)$ (all) &0.0973 &0.0972 &0.0869 \\
$wR_{F }$ $(m = 2)$ (all) &0.3748 &0.3721 &0.1319 \\
$wR_{F }$ all (all) &0.0879 &0.0877 &0.0738\\
GoF (obs/all) &1.69/1.23 &1.69/1.23 &1.52/1.03 \\
$\Delta\rho_{min}$, $\Delta\rho_{max}$ (e \AA{}$^{-3}$) &
 -14.56, 14.53 & -14.52, 14.69 &-8.73, 6.14 \\
\hline
\end{tabular}
\end{table}

\subsubsection{\label{sec:ho2ir3si5_model_a}%
Model A}

Here, up to second-order harmonics have been
applied to holmium and iridium atoms, whereas
silicon atoms have only first-order harmonics
(Eq. S6 in \cite{ho2ir3si5suppmat2022a}).
This resulted in fewer parameters
and minimal changes to the results
as compared to models B and C
(Table \ref{tab:ho2ir3si5_refmod_compare}).
This model has been selected to describe the crystal
structure in the CDW phase.

\subsubsection{\label{sec:ho2ir3si5_model_b}%
Model B}

In this model up to second-order displacement
modulation parameters have been applied to all
atoms and then refined.
The refinement resulted in a fit to the SXRD data
that is almost identical to that of Model A.
However, the second-order modulation amplitudes
of the silicon atoms appear to have large
standard uncertainties (s.u.'s).
This is probably due to the poor scattering power
of silicon as compared to holmium or iridium.
Together, these features imply that the
additional 30 parameters in model B do not
lead to a significant improvement and model
B is not selected.

\subsubsection{\label{sec:ho2ir3si5_model_c}%
Model C}

Model C builds upon model A and model B
in order to try solve the issue with the
high value of $R_{F}$ for the second-order
satellites.
Up to third-order harmonics are used for the
modulation functions of the holmium and iridium
atoms, while employing only first-order
harmonics for silicon atoms, as higher-order
harmonics have been discarded for Si on the
basis of models A and B.
As a rule of thumb the highest harmonics in the
modulation wave should not exceed the highest
order of observed satellite reflections,
the latter which is two in the present SXRD experiment.
At first glance from Table \ref{tab:ho2ir3si5_refmod_compare}
we see a huge improvement of the fit to the
second-order satellites upon introduction of
third-order harmonics.
However, all these parameters have a s.u.
that is larger than the refined value itself.
The latter indicate that the system would have
third-order satellite reflections with
intensities larger than its first- and second-order
satellites.
As we did not observe third-order satellites,
the refined values of the third-order harmonics
appear to be unreliable, and model A is chosen
in favor of model C.

\subsection{\label{sec:ho2ir3si5_location_cdw}%
Location of the CDW}

\begin{figure}%[H]
\includegraphics[width=160mm]{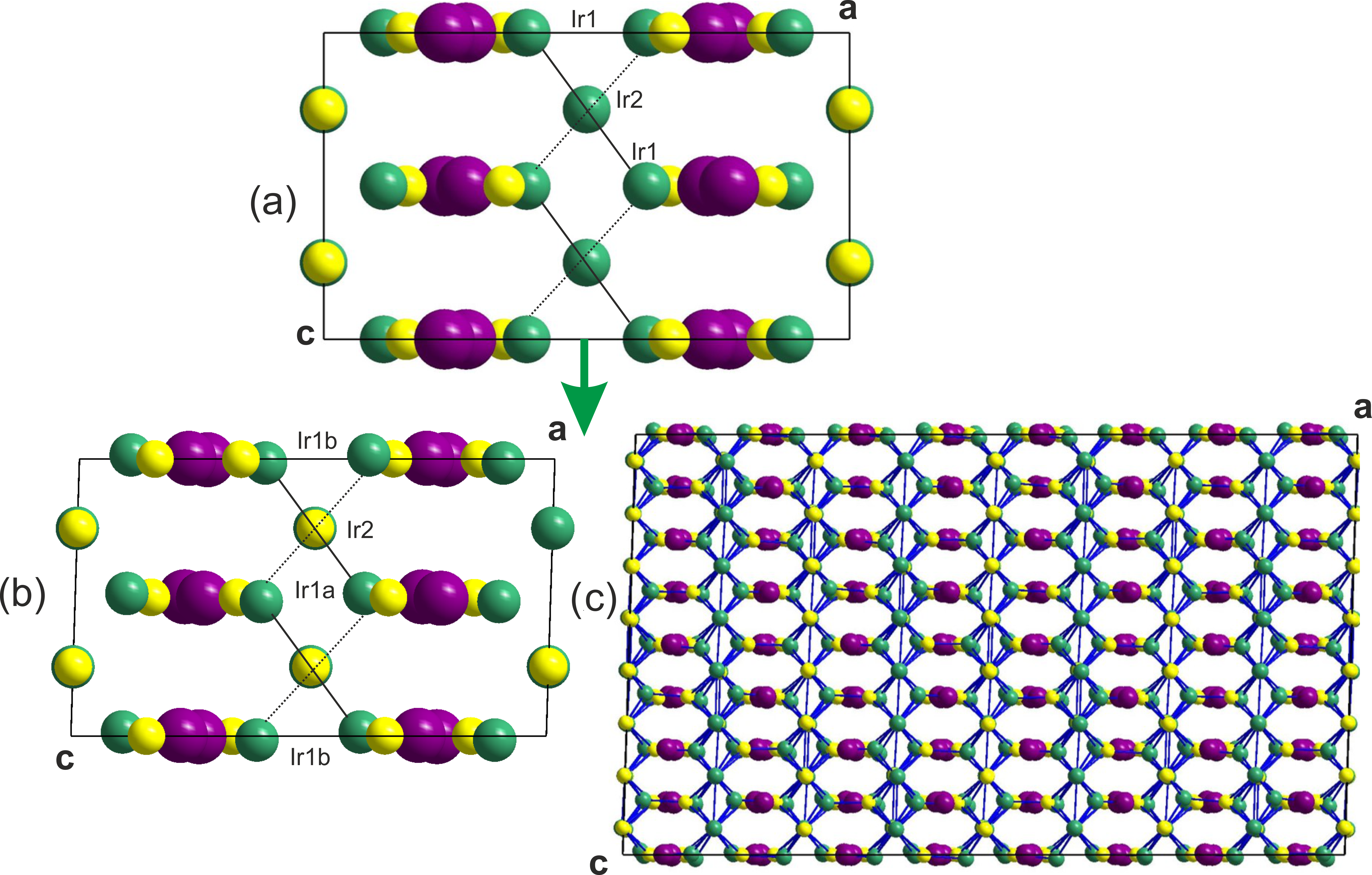}
\caption{\label{fig:ho2ir3si5_cell}%
Projection onto the $(\mathbf{a},\mathbf{c})$-plane
of the crystal structures of Ho$_{2}$Ir$_{3}$Si$_{5}$ for
(a) 200 K and (b,c) 70 K.
Panel (c) shows the $4\times 2 \times 4$
superstructure approximation.
Large purple spheres correspond to Ho atoms;
green spheres of intermediate size correspond to Ir atoms;
small yellow spheres are Si atoms.
Dashed lines give the distances in the basic structure,
with values of $3.390\,(3)$ and $3.728\,(3)$ \AA{}.}
\end{figure}

Table S8 in the Supporting Information \cite{ho2ir3si5suppmat2022a}
shows the atomic coordinates at 200 K and 70 K (warming).
The six crystallographically independent atoms,
Ho1, Ir1, Ir2, Si1, Si2, and Si3,
of the $Ibam$ structure at 200 K
split into the
Ho1a, Ho1b, Ir1a, Ir1b, Ir2, Si1, Si2a, Si2b, Si3a, Si3b
atoms of the triclinic structure at 70 K.
Figure \ref{fig:ho2ir3si5_cell} shows the crystal
structure projected onto the \textbf{a-c} plane
at 200 K and 70 K (average structure at 70 K).
Figure \ref{fig:ho2ir3si5_cell}(b) appears skew
compared to Fig. \ref{fig:ho2ir3si5_cell}(a) due
to $\beta > 90$ deg in the CDW phase.
As the modulation wave vector
is close to (1/4, 1/2, 1/4),
Fig. \ref{fig:ho2ir3si5_cell}(c) shows a
$4\times 2 \times 4$ superstructure approximation,
where one can see zigzag chains of Ir1a-Ir1b
along \textbf{c}.
Tables S9 and S10 in the
Supporting Information \cite{ho2ir3si5suppmat2022a}
show the refined modulation amplitudes.

\begin{figure}%[H]
\includegraphics[width=80mm]{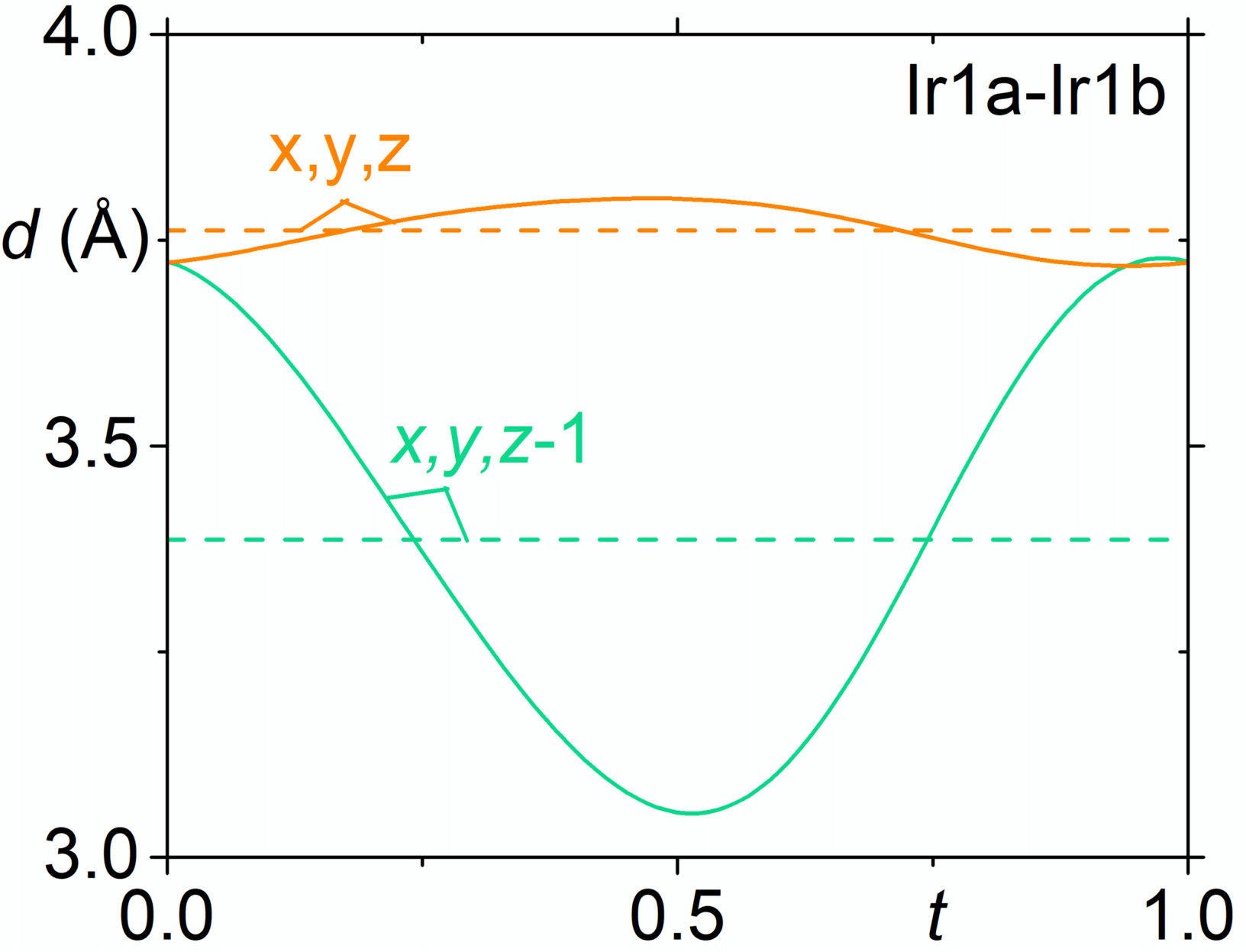}
\caption{\label{fig:ho2ir3si5_tplot}%
$t$-Plot of the interatomic distances between atoms
Ir1a and Ir1b $(x,y,z)$ and between Ir1a and
Ir1b at $(x, y, z-1)$ for the crystal in the CDW
phase at $T=70$ K.}
\end{figure}

Analysis of the modulation
amplitudes and distances revealed that Ho$_2$Ir$_3$Si$_5$ follows
a similar pattern as $R_2$Ir$_3$Si$_5$ (Lu, Er),
\cite{ramakrishnan2021a,ramakrishnan2020a}
such that the CDW resides on zigzag chains
of Ir1a-Ir1b atoms,
as they have the shortest metal-metal
distances and exhibit the largest modulation of these
distances (Fig. \ref{fig:ho2ir3si5_tplot}).
%
%SvS-removed the following sentence
%A $t$-movie is also created using the
%MoleCoolQt software \cite{hubschle2011a}
%to exhibit this feature (citation needed).
%
The lattice distortion results in alternating
short and long Ir1a--Ir1b distances, where the
shorter distance is the most affected by the
modulation.
The formation of dimers on the Ir1a-Ir1b
zigzag chains along \textbf{c}
is responsible for the formation of the CDW.

Table \ref{tab:lu2ir3si5_compare} provides a
comparison of essential structural parameters
between the three compounds $R_2$Ir$_3$Si$_5$.
Our earlier proposal, that $T_{CDW}$ would be
inversely proportional to the atomic radius of
the rare earth element, is not confirmed by
Ho$_2$Ir$_3$Si$_5$.
The atomic radius of Ho is comparable to that
of Er, however $T_{CDW}$ of Ho$_2$Ir$_3$Si$_5$
is much lower than that of Er$_2$Ir$_3$Si$_5$.
The variation of distances Ir1a--Ir1b is more
or less similar in all three compounds.
\begin{table}%%[H]
%\scriptsize
%\small
%\centering
\caption{\label{tab:lu2ir3si5_compare}%
Crystal data of the CDW phases of
Lu$_{2}$Ir$_{3}$Si$_{5}$ \protect\cite{sangeetha2015a, ramakrishnan2021a},
Er$_{2}$Ir$_{3}$Si$_{5}$ \protect\cite{ramakrishnan2020a}
and Ho$_{2}$Ir$_{3}$Si$_{5}$ (present results).
}
\begin{tabular}{c c c c}
\hline
Compound & Lu$_{2}$Ir$_{3}$Si$_{5}$ & Er$_{2}$Ir$_{3}$Si$_{5}$ & Ho$_{2}$Ir$_{3}$Si$_{5}$  \\
\parbox[c]{24mm}{Atomic radius of $R$ (\AA{}) \cite{clementi1967a}}
&  2.17 &  2.26 & 2.26 \\
$T_{\mathrm{CDW}}$ (K) & 202--231 & 150--166 & 90--130 \\
$T$ (K) & 60 & 75  & 70  \\
$a$ (\AA{}) & 9.8182(3)  &  9.8494(3) &  9.8356(5) \\
$b$ (\AA{}) & 11.4093(3) &  11.4863(3) &  11.4902(4) \\
$c$ (\AA{}) & 5.6835(2)  &  5.7268(2)  &  5.7304(3) \\
$\alpha$ (deg) & 90.001(2)  & 90.079(1) &  89.983(3) \\
$\beta$ (deg)  & 91.945(2)  & 91.695(2)& 91.772(2) \\
$\gamma$ (deg) & 90.018(2)  & 90.051(1)&  89.975(1) \\
$V$ (\AA{}$^3$)& 636.34(3) & 647.60(5) &  647.32(5) \\
$q_x$ & 0.2499(3) & 0.2495(2)& 0.2494(2) \\
$q_y$ & 0.4843(4) & 0.4973(1)& 0.4978(2) \\
$q_z$ & 0.2386(2) & 0.2483(1)& 0.2488(2) \\
\multicolumn{2}{l}{Distance Ir1a--Ir1b\textsuperscript{\emph{a}}} \\
max (\AA{}) & 3.801(1) & 3.818(2) & 3.801(3) \\
min (\AA{}) & 3.711(1) & 3.714(2) & 3.719(3) \\
avg (\AA{}) & 3.755(1) & 3.764(2) & 3.763(3) \\
\multicolumn{2}{l}{Distance Ir1a--Ir1b\textsuperscript{\emph{b}}} \\
max (\AA{}) & 3.761(1) & 3.782(2) & 3.728(3) \\
min (\AA{}) & 3.002(1) & 3.008(2) & 3.053(3) \\
avg (\AA{}) & 3.385(1) & 3.398(2) & 3.390(3) \\
\hline
\end{tabular}\\
\textsuperscript{\emph{a}}Symmetry code for Ir1b $(x,y,z)$; given are the
maximum (max), minimum (min) and average (avg) distances.
\textsuperscript{\emph{b}}Symmetry code for Ir1b $(x,y,z-1)$.
\end{table}

\subsection{\label{sec:ho2ir3si5_electronic_structure}%
Electronic structure and phonons}

\begin{figure}%[H]
\includegraphics[width=0.95\textwidth]{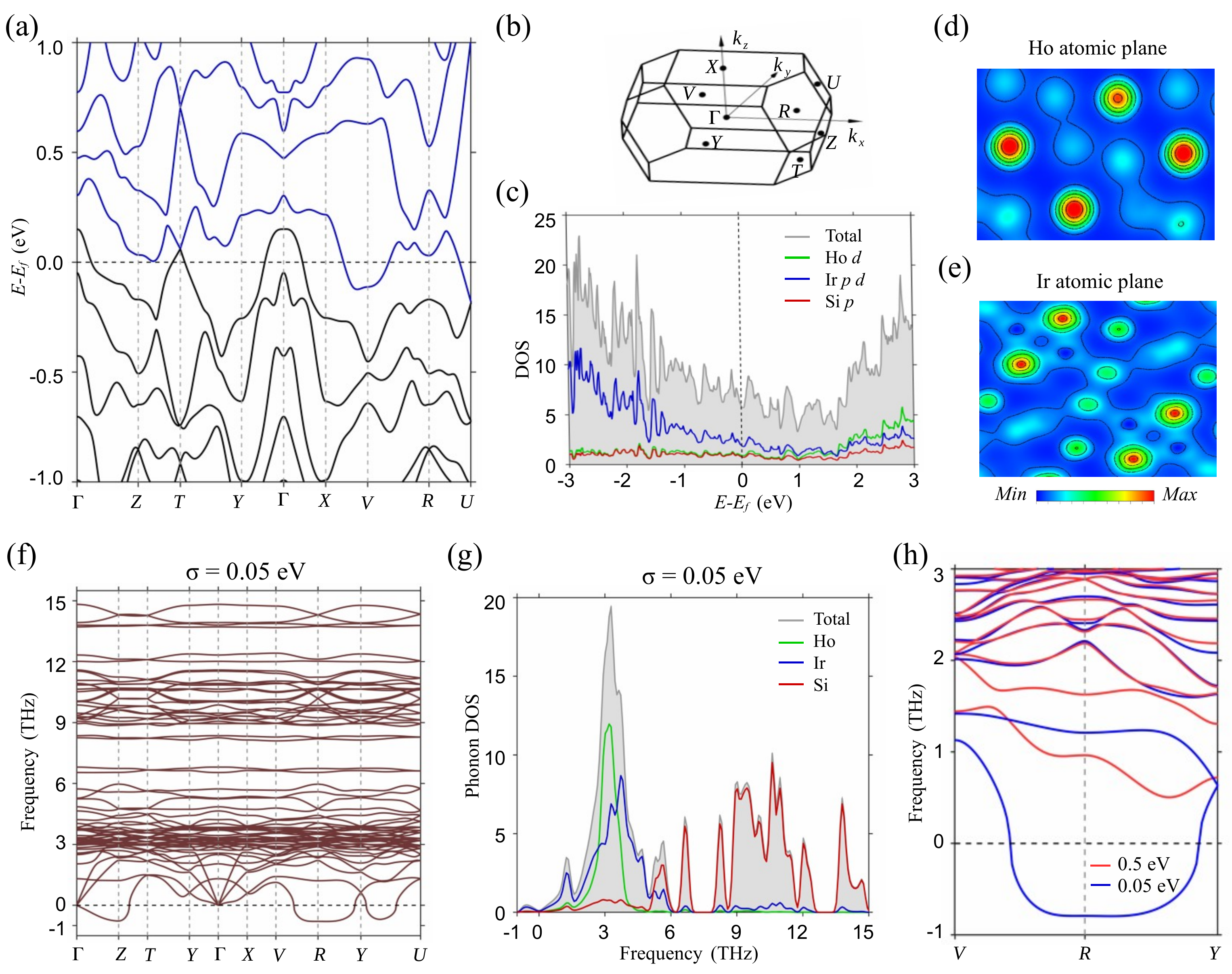}
\caption{\label{fig:ho2ir3si5_band}%
(a) Bulk electronic band structure of the orthorhombic
phase of Ho$_2$Ir$_3$Si$_5$ along various high-symmetry
directions in the primitive Brillouin zone.
(b) Brillouin zone with high-symmetry points.
(c) Total (highlighted grey color) and orbital
projected density of states (green, red, and
blue lines).
(d) and (e) Electronic charge density distributions
on planes containing Ho and Ir atoms, respectively.
(f) Phonon band structure of the orthorhombic phase,
as calculated with smearing parameter $\sigma = 0.05$ eV.
(g) Phonon density of states.
The imaginary frequencies are represented by
negative values and are mainly associated with Ir atoms.
(h) Expanded view of the phonon band structure along
the $V$--$R$--$Y$ directions.
The absence of imaginary frequencies for smearing
parameter $\sigma = 0.5$ eV indicates the stability
of the orthorhombic structure towards higher temperatures.
}
\end{figure}

The band structure of Ho$_2$Ir$_3$Si$_5$
is shown in Fig.~\ref{fig:ho2ir3si5_band}(a)
for several high-symmetry directions in the
primitive BZ given in Fig.~\ref{fig:ho2ir3si5_band}(b).
It is metallic such that both
electron and hole pockets exist at the Fermi level.
To resolve the contributions of electronic
states near the Fermi energy, we show the
atomic orbital projected density of states (PDOS)
in Fig.~\ref{fig:ho2ir3si5_band}(c).
The Ir states are dominant near $E_F$, whereas
the Ho and Si states have lower weight than those
of Ir atoms.
To understand the charge density distributions
of these atoms, we have shown the charge density
on two different planes in the
primitive unit cell, which contain Ho and Ir atoms,
respectively [Figs.~\ref{fig:ho2ir3si5_band}(d)
and~\ref{fig:ho2ir3si5_band}(e)].
The spherical charge distribution of these atoms
is reminiscent of the metallic type ionic environment
and suggests that these atoms are more likely to
undergo CDW modulations to find a low energy state.\cite{Bartl1979a}

Figure~\ref{fig:ho2ir3si5_band}(f) shows the phonon
spectrum along various high-symmetry
directions in the BZ.
The system is dynamically
unstable with Kohn-type~\cite{kohn1959image}
soft modes at $Z$, $R$ and in between
$Y$ and $U$ points.
The phonon PDOS [Fig.~\ref{fig:ho2ir3si5_band}(g)]
shows that Ho and Ir atoms contribute to the
low-frequency phonon modes, whereas high-frequency
phonon modes are dominated by Si atoms.
The imaginary frequencies are inherent to the Ir atoms,
confirming that the CDW is associated mainly
with the Ir atoms.
The lowest value of negative frequency is found
near the reciprocal point (0.25, 0.50, 0.25),
which is consistent with the incommensurate
wave vector $\mathbf{q}$ found in the SXRD experiment.
Figure~\ref{fig:ho2ir3si5_band}(h) demonstrates
the evolution of the soft phonon mode as a function
of the smearing parameter $\sigma$,
which represents the electronic temperature in our
calculations.
The soft phonon mode disappears upon increasing
the smearing parameter $\sigma$ from 0.05 eV to 0.5 eV.
This particular dependence of the phonon frequencies
on the smearing parameter (electronic temperature)
indicates that the orthorhombic structure is stable
only at higher temperatures as seen in our
experiments.

\subsection{\label{sec:ho2ir3si5_electrical_resistivity}%
Electrical resistivity}

\begin{figure}%[H]
\includegraphics[width=80mm]{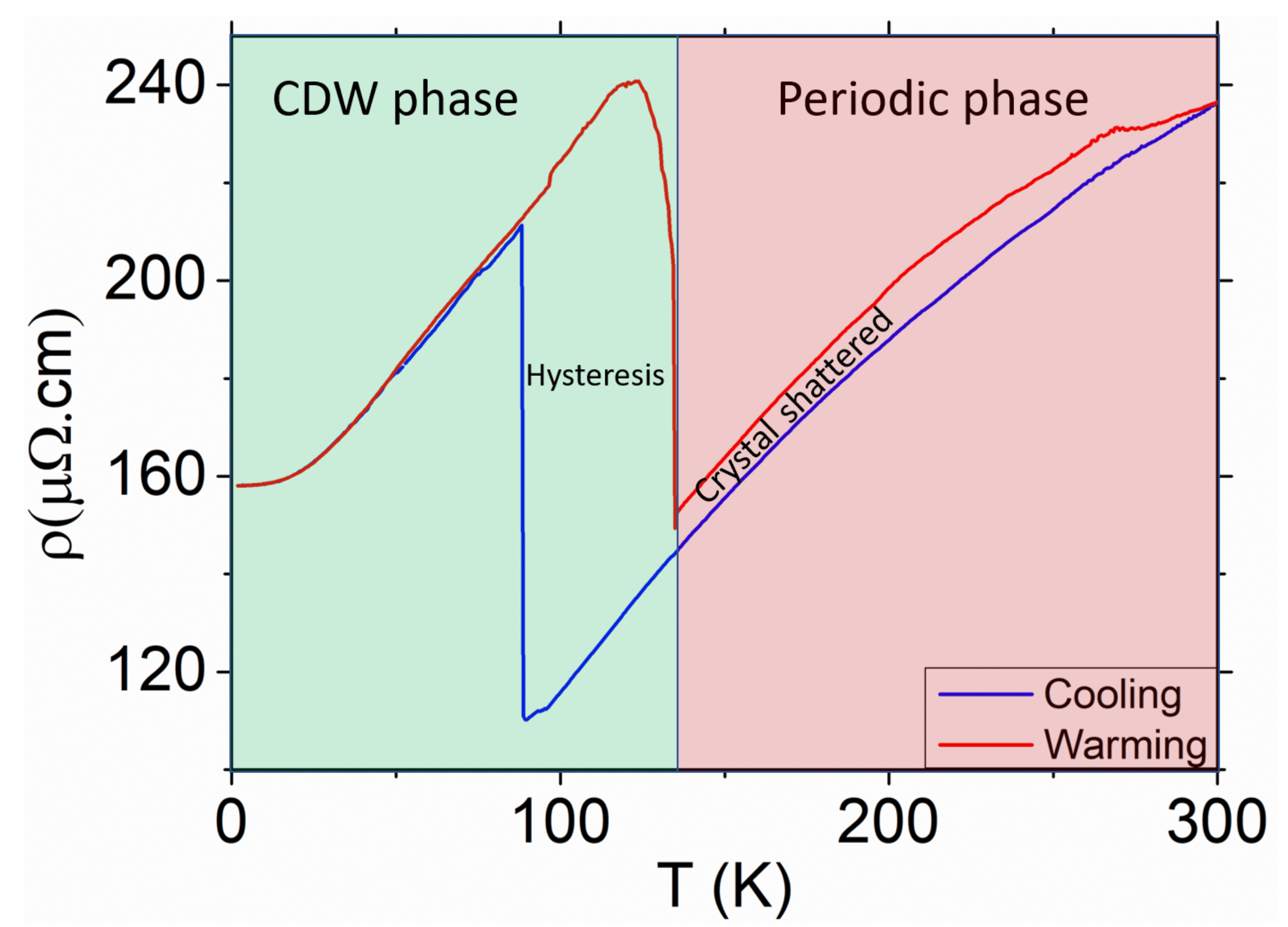}
\caption{\label{fig:ho2ir3si5_resi}%
Temperature dependence of the electrical
resistivity $\rho(T)$ of Ho$_2$Ir$_3$Si$_5$.
The pink and green regions indicate the periodic
and CDW phases, respectively.
A large hysteresis of 40 K is clearly seen.
The crystal shattered on heating to above 130 K.}
\end{figure}

From Fig. \ref{fig:ho2ir3si5_resi} we infer that,
upon cooling the crystal, there is a sharp upward
turn of the electrical resistivity at 90 K, which
signifies the opening of a gap in the electronic
density of states over a major fraction of the
Fermi surface, in agreement with the formation of
a charge density wave (CDW) at this temperature.
The metallic nature of the crystal after the
CDW transition indicates a partial opening of
the gap at the Fermi surface.
Upon heating, $\rho(T)$ shows a sharp anomaly
at 130 K, thereby establishing a huge hysteresis
of 90--130 K.
This large hysteresis and the sharpness of the
transition signify that it is a first-order
transition much like the one seen
in $R_2$Ir$_3$Si$_5$ ($R$ = Lu, Er)
\cite{sangeetha2015a, ramakrishnan2020a}.
Such a large hysteresis is uncommon
for CDW transitions.
Especially for incommensurate CDWs the transition
usually is of second order, \textit{e.g.}
as found in the canonical CDW system
NbSe$_3$ \cite{tomi1981a}.

All measurements were made using virgin samples.
After cooling and then heating through the phase
transition,
micro-cracks develop due to the strain induced
by the transition.
This feature is visible in the electrical
resistivity through a lower resistivity of
virgin samples than of samples that have gone
through a cooling/heating cycle
(Fig. \ref{fig:ho2ir3si5_resi}).
Similar behavior has been observed for
ceramic material of $R_2$Ir$_3$Si$_5$
($R$ = Lu, Er) \cite{singhy2004a, singhy2005a}.
It was explained
by variations of pinning of the CDW in those systems.
However, it is
possible that the large lattice distortions
accompanying the CDW
transition could be responsible for the
formation of micro-cracks.
This leads to a
reduction of sizes of mosaic blocks, and an increase in texture,
which, in turn, would cause an increase of the electrical resistance
at each thermal cycle.
The presence of micro-cracks is
supported by the observation that the transition temperatures
and hysteresis are the same in each thermal cycle.
On the other hand,
the increase of pinning centers would lead to a lowering of the CDW
transition temperature, which is not observed here.
Just as in the case of $R_2$Ir$_3$Si$_5$ ($R$ = Lu, Er)
\cite{ramakrishnan2021a, ramakrishnan2020a},
we observe that the electrical resistivity exhibits
a $T^2$ dependence below $T$ = 50 K, which is up to much
higher temperatures than $\sim$10 K for a Fermi liquid,
implying dominant contributions from short-range
magnetic fluctuations of Ho spins in the absence of
long-range magnetic ordering.

\subsection{\label{sec:ho2ir3si5_magnetic_susceptibility}%
Magnetic susceptibility}

\begin{figure}%[H]
\includegraphics[width=80mm,keepaspectratio]{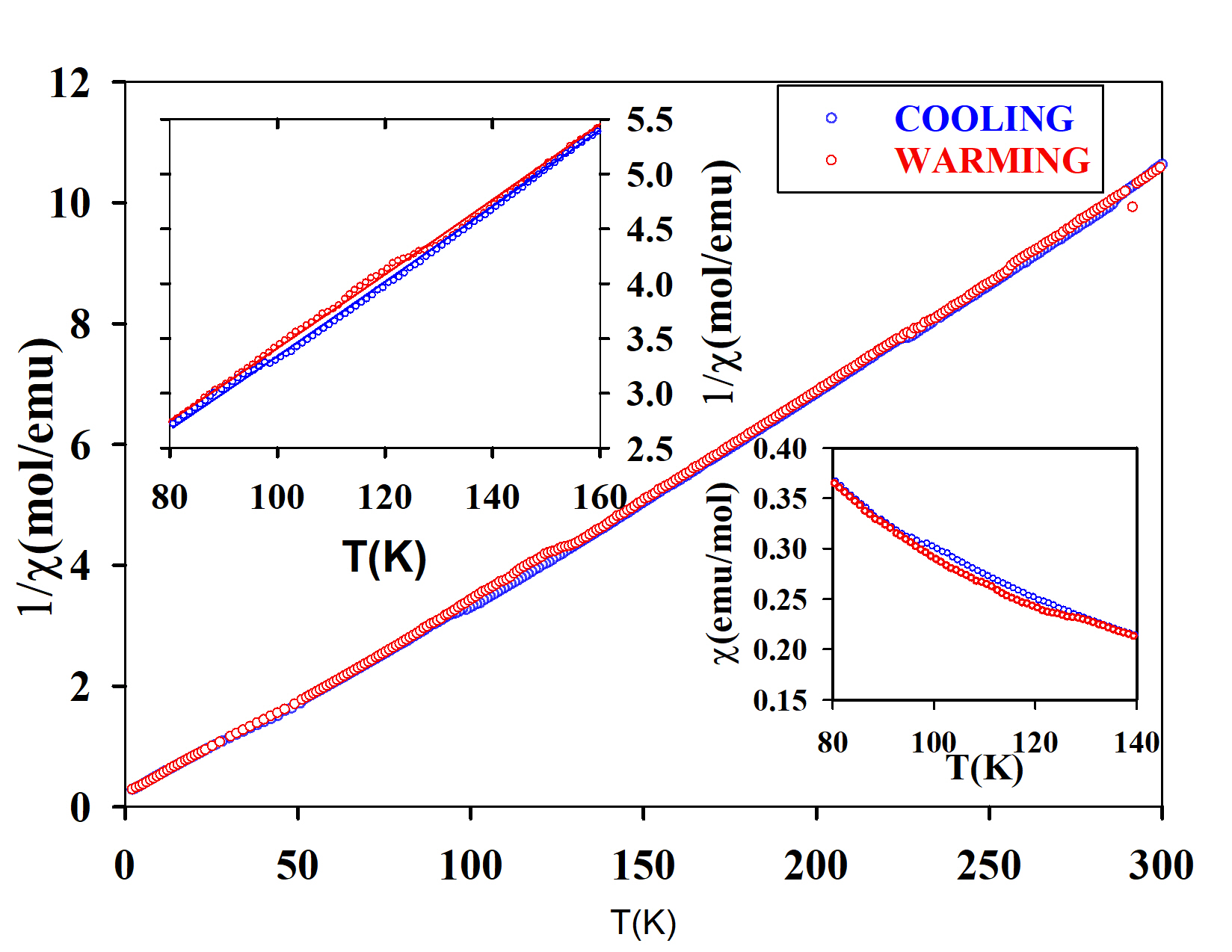}
\caption{\label{fig:ho2ir3si5_suscep}%
Temperature dependence of the inverse magnetic
susceptibility ($1/\chi(T)$) of Ho$_2$Ir$_3$Si$_5$.
The upper inset shows an expanded view of $1/\chi(T)$
versus temperature.
The lower inset shows an expanded view $\chi(T)$
versus temperature.}
\end{figure}

The temperature-dependent magnetic susceptibility,
$\chi(T)$, of Ho$_2$Ir$_3$Si$_5$ was measured on
a single crystal from the same rod as used for the
other bulk measurements.
Measurements were performed with a commercial
superconducting quantum interference device
(SQUID) magnetometer (MPMS 7, Quantum Design, USA)
employing a magnetic field of 0.1 T
along $\mathbf{c}$, during cooling
and heating between 2 and 300 K
(Fig. \ref{fig:ho2ir3si5_suscep}).
The most interesting feature of the
temperature-dependent
magnetic susceptibility is the noticeable increase
of $\chi$(T) at 90(1) K upon cooling through
the CDW transition.
The observed effect cannot be explained by
a change of Pauli susceptibility at the
transition, since the estimated magnitude
of the Pauli susceptibility is at least two orders
of magnitude smaller than the observed change in
the susceptibility.
Furthermore, one would expect a smaller Pauli
susceptibility in the CDW phase, whereas we
observe an increase of the
susceptibility when cooling through the transition.
The same transition is found at 130(1) K
upon heating.
The observed hysteresis is in good agreement with
the hysteresis in the electrical resistivity and
DSC measurements.
A Curie-Weiss fit to the data at 140--300 K
results in a Curie constant of $C$ = 31.78(1) emu/mol K
and an antiferromagnetic Weiss temperature of
$\theta$ = -2.6(1) K.
A Curie-Weiss fit to the low-temperature paramagnetic
regime 40--90 K results in $C$ = 32.4(1) emu/mol K
and $\theta$ = -6.3(1) K.
The different Curie constants correspond to different
effective magnetic moments on Ho$^{3+}$
of 11.27 $\mu_B$ and 11.38 $\mu_B$, respectively.
These values are slightly higher than the free ion
magnetic moment of Ho$^{3+}$ ions.
%free ion value of Ho3+ is 10.60 $\mu_B$.

So we see that Ho$_2$Ir$_3$Si$_5$ joins our
earlier studied Er$_2$Ir$_3$Si$_5$ \cite{ramakrishnan2020a}
as yet another exceptional case in showing
an effect of the CDW transition on the magnetic susceptibility.
Usually, compounds containing magnetic
rare-earth elements do not show any anomaly
in the paramagnetic susceptibility at the
high-temperature CDW transitions.
This is true for many of the magnetic CDW
compounds which are mentioned in the introduction.
For example, the paramagnetic susceptibility
of Ho$_5$Ir$_4$Si$_{10}$ does not show any anomalies
at its CDW transition \cite{yanghd1991c,ghoshk1993}.
%For this compound, the Curie-Weiss fit resulted
%in a magnetic moment on Ho$^{3+}$ of 11.3 $\mu_B$
%\cite{ramakrishnan2017a}.
%Furthermore, the interaction
%between magnetic moments appears to be much weaker
%in Ho$_5$Ir$_4$Si$_{10}$
%(fitted value $\theta$ = 3.0 K) than in Ho$_2$Ir$_3$Si$_5$.
The coexistence of antiferromagnetic order
($T_N$ = 2 K) and CDW in Ho$_5$Ir$_4$Si$_{10}$
might be related to the presence of weakly
coupled $4f$ electrons of Ho$^{3+}$
ions \cite{ramakrishnan2017a},
while for Ho$_2$Ir$_3$Si$_5$ the reduced
magnitude of the magnetic moments and
strong coupling suggest influence of the $4f$
electrons in the CDW transition and vice versa.
It is not clear, whether
the small but distinct change of Ho$^{3+}$
moment (11.27 $\mu_B$ vs 11.38 $\mu_B$)
across the CDW transition in a single crystal
of Ho$_2$Ir$_3$Si$_5$ could be responsible for
the absence of magnetic ordering of Ho$^{3+}$
moments in the crystal down to 2 K.

\subsection{\label{sec:ho2ir3si5_specific_heat}%
Specific heat}

\begin{figure}%[H]
\includegraphics[width=80mm,keepaspectratio]{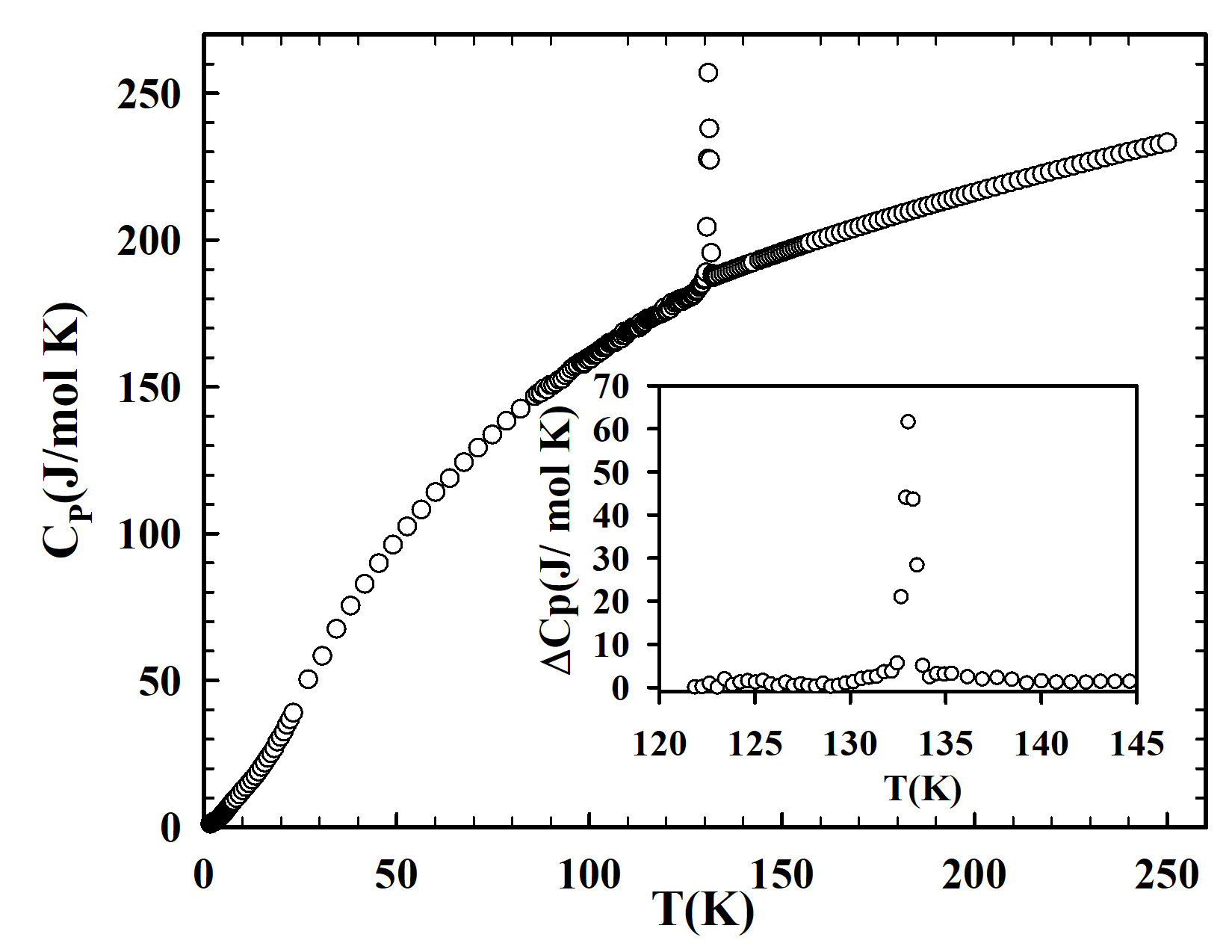}
\caption{\label{fig:ho2ir3si5_specific_heat}%
Temperature dependence of the specific heat
$C_p$ from 2 to 250 K using PPMS.
The inset provides an enlarged view of the CDW
transition where $\Delta C_p$ = 60.5(1) J/(mol K).}
\end{figure}
\begin{figure}%[H]
\includegraphics[width=80mm]{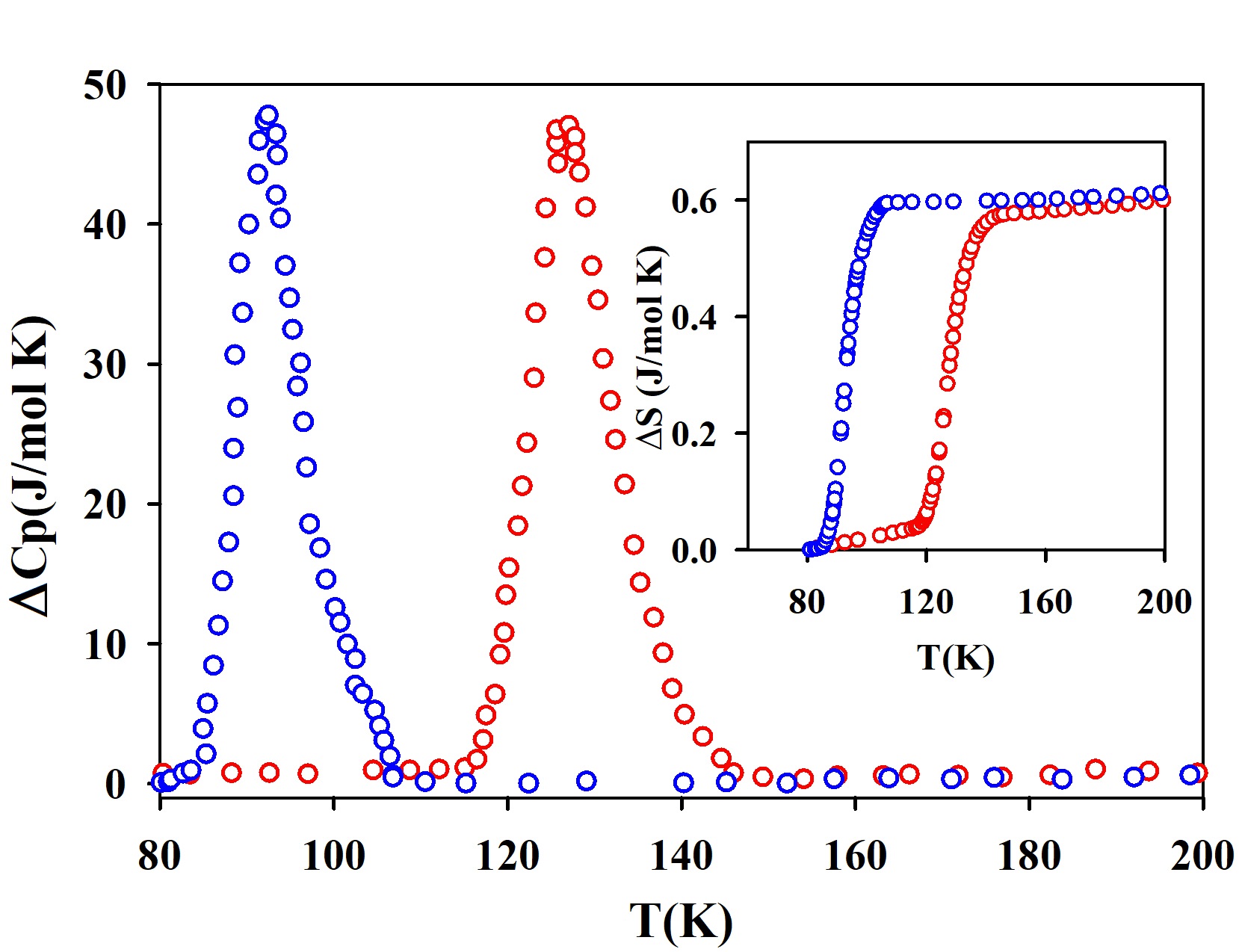}
\caption{\label{fig:ho2ir3si5_dsc}%
Temperature dependence of the excess specific heat
$\Delta C_p$,
obtained by subtracting a smooth baseline from the
DSC signal.
Blue and red circles refer to cooling and warming data.
Clear peaks are observed in both cooling and
warming with a hysteresis of about 40 K.}
\end{figure}

The specific heat ($C_p(T)$) of Ho$_2$Ir$_3$Si$_5$
was measured by the thermal relaxation method,
using a physical property measuring system
(PPMS, Quantum Design, USA).
Data obtained during heating of a
single crystal of 10.5 mg from 2 to 250 K exhibit a sharp
peak at the temperature of the CDW transition at 131.1 (1) K
(Fig. \ref{fig:ho2ir3si5_specific_heat}).
We are unable to find a peak in $C_p(T)$ while cooling
the crystal.
Similar behavior was noted earlier for a crystal of
Lu$_2$Ir$_3$Si$_5$ \cite{sangeetha2015a}.
This feature could be the result of the specific
method of measurement employed in the PPMS instrument.
In a second experiment, differential scanning calorimetry (DSC)
was measured from 80 to 150 K on the same single crystal
of Ho$_2$Ir$_3$Si$_5$ (Fig. \ref{fig:ho2ir3si5_dsc}).
The DSC data show clear peaks at different temperatures
in the heating and cooling runs,
that appear at similar temperatures as are found
for the electrical resistivity and magnetic susceptibility,
and thus confirm the first-order character of the
phase transition.
The peaks appear of similar width as for
Lu$_2$Ir$_3$Si$_5$ \cite{sangeetha2015a},
while the DSC peaks are much sharper for
Er$_2$Ir$_3$Si$_5$ \cite{ramakrishnan2020a}.
Since other features, like $C_p(T)$ and $\rho(T)$,
exhibit very sharp anomalies at the CDW transition,
the broadened features in the DSC signal might
be related to the rate of change of temperature
in this experiment in conjunction with the sluggish
character of the transition.

The lattice contribution to the specific heat was determined
from a fit to the data far away from the transition.
Subtraction
of the lattice contribution resulted in $\Delta C_p(T)$
(inset in Fig. \ref{fig:ho2ir3si5_specific_heat}).
A similar value was obtained from the DSC measurement
(Fig. \ref{fig:ho2ir3si5_dsc}).
The change of entropy at the transition $\Delta S$
has been determined by integration of
$\tfrac{\Delta C_p(T)}{T}$ over temperature.
The values of $\Delta C_p$ = 60.5(1) J/(mol K)
and $\Delta S$ = 0.6 J/mol
are comparable to transition entropies
for $R_5$Ir$_4$Si$_{10}$
and $R_2$Ir$_3$Si$_5$ \cite{ramakrishnan2017a,sangeetha2015a}.
However, they are much larger than obtained
for conventional CDW systems, such as
K$_{0.3}$MoO$_{3}$ ($\Delta C_p$(max) = 8 J/(mol K);
$\Delta S$ = 0.18R)\cite{Bartl1979a} and
NbSe$_3$ ($\Delta C_p$(max) =9 J/(mol K);
$\Delta S$ = 0.08R)\cite{tomi1981a}.
The specific heat anomaly indicates a much
sharper transition for single crystals
of Ho$_2$Ir$_3$Si$_5$ than for conventional CDW
systems, which is in agreement with the first-order
character of the transition
as deduced from
resistivity and magnetic susceptibility data.

\section{\label{sec:ho2ir3si5_conclusions}%
Conclusions}

We have established an incommensurately
modulated crystal structure of the CDW phase
of Ho$_2$Ir$_3$Si$_5$.
The incommensurate modulation
is accompanied by a strong lattice distortion,
both of which are important for the
modulation of interatomic distances on
zigzag chains of iridium atoms along $\mathbf{c}$.
This is in accordance with the CDW being
supported by these zigzag chains.
Similar to the case of Er$_2$Ir$_3$Si$_5$,
the rare earth atoms
are not directly involved in the CDW formation.
The occurrence of a large lattice distortion
accounts for the sluggish character
and large hysteresis of the transition,
as they are apparent
in the temperature dependencies of the electrical resistivity,
magnetic susceptibility and specific heat.
Another unique feature of the compounds
$R_2$Ir$_3$Si$_5$ ($R$ = Lu, Er and Ho)
is the extreme sensitivity of the phase
transitions to crystalline order.
The present single crystals of high
perfection undergo a CDW transition,
while magnetic order is suppressed
down to at least 1.5 K.
It is worthwhile to point out that the
previous experiments on polycrystalline material
did not observe the CDW transition,
while magnetic order appeared
below $T_N$ = 5.1 K \cite{singhy2004a, singhy2005a}.
The present results on Ho$_2$Ir$_3$Si$_5$
do not confirm the idea that $T_{CDW}$ would
scale with the atomic radius of the rare earth
element.
While the size of Ho is equal to
that of Er, $T_{CDW}$ is much lower for
Ho$_2$Ir$_3$Si$_5$ than for Er$_2$Ir$_3$Si$_5$.
This absence of a correlation in the
series $R_2$Ir$_3$Si$_5$ is in contrast to
the series of isostructural rare-earth compounds
$R_5$Ir$_4$Si$_{10}$  \cite{ramakrishnan2017a},
where $T_{CDW}$ increases with increasing
size of the rare earth element $R$.

The SXRD data have revealed that the
structure of Ho$_2$Ir$_3$Si$_5$ is different
from that of Lu$_2$Ir$_3$Si$_5$ and
Er$_2$Ir$_3$Si$_5$.
Although the symmetries are the same for
all three compounds,
the presence of second-order superlattice
reflections in the CDW phase of Ho$_2$Ir$_3$Si$_5$
has added more complexity to it,
requiring modulation functions with up to
second-order harmonics (model A in
Section \ref{sec:ho2ir3si5_cdw_structure}).

From physical property measurements,
a huge hysteresis is found of about 40 K,
with the transition proceeding at
90 K (cooling) and 130 K (warming).
The transition implies severe structural
distortions, such that sometimes the crystal
shatters.
This is visible in the electrical resistivity,
which is higher after cycling though the
transition than before (Fig. \ref{fig:ho2ir3si5_resi}).
The effective magnetic moment of Ho is found
to change at the CDW transition,
from 11.38 $\mu_B$ to 11.27 $\mu_B$
(Fig. \ref{fig:ho2ir3si5_suscep}).
This kind of coupling between CDW and magnetism
is rare.
It has only been observed in isostructural
Er$_2$Ir$_3$Si$_5$.
This feature can probably be attributed to the
participation of $4f$ orbitals of Ho or Er
in states near the Fermi level that are involved
in CDW ordering.
The fact the effective moment of the rare earth
is influenced by the CDW transition suggests
a competition between CDW and magnetic order,
with both interactions presumedly employing
the same part of the Fermi surface.
In order to clearly understand the mechanism
for this unusual nature one would probably need
microscopic probes of magnetism,
such as elastic and inelastic neutron scattering.

\begin{acknowledgement}
We acknowledge DESY (Hamburg, Germany),
a member of the Helmholtz Association HGF,
for the provision of experimental facilities.
Parts of this research were carried out at
PETRA III, using beamline P24.
J.-K. Bao acknowledges financial support from the
Alexander-von-Humboldt foundation.
The work at TIFR Mumbai was supported by the
Department of Atomic Energy of the Government
of India under Project No. 12-R\&D-TFR-5.10-0100.
This research has been funded by the Deutsche
Forschungsgemeinschaft
(DFG; German Research Foundation)--406658237.
\end{acknowledgement}


\begin{suppinfo}
Details of the SXRD data collection,
data processing and structural analysis.
This material is available free of charge
\textit{via} the Internet at
\texttt{http://pubs.acs.org}.
%
%The following files are available free of charge.
%\begin{itemize}
%\item SXRD supporting information: Details of SXRD data
%collection, processing and analysis.
%\item t-movie: t-movie to illustrate how the modulation
%affects the chains of iridium atoms.
%\end{itemize}
%
\end{suppinfo}

%\bibliography{Ho_main}
\providecommand{\latin}[1]{#1}
\makeatletter
\providecommand{\doi}
  {\begingroup\let\do\@makeother\dospecials
  \catcode`\{=1 \catcode`\}=2 \doi@aux}
\providecommand{\doi@aux}[1]{\endgroup\texttt{#1}}
\makeatother
\providecommand*\mcitethebibliography{\thebibliography}
\csname @ifundefined\endcsname{endmcitethebibliography}
  {\let\endmcitethebibliography\endthebibliography}{}

\end{document}